\documentclass[10pt]{article} 
\usepackage{amsmath}
\usepackage{amsfonts}
\def\non{ \nonumber }
\begin{document} 
\def\a{\alpha}
\def\b{\beta}
\def\({\left(}
\def\){\right)}
\def\[{\left[}
\def\]{\right]}
\def\J{J(X)}
\def\X{\Theta}
\def\lar{\longrightarrow}
\def\llar{\longleftarrow}
\def\ot{\otimes}
\def\S{X}
\def\ra{\rightarrow}
\def\afx{X_{\text{aff}}}
\def\mbc{\mathbb{C}}

\rightline{LPTHE-0002}
\vskip 1cm
\centerline{\LARGE Cohomologies of Affine Jacobi Varieties  }
\bigskip
\centerline{\LARGE  and Integrable Systems .}
\vskip 2cm
\centerline{\large 
A. Nakayashiki ${}^a$ and F.A. Smirnov ${}^b$
\footnote[0]{Membre du CNRS}}
\vskip1cm
\centerline{ ${}^a$ Graduate School of
Mathematics, Kyushu University} 
\centerline{Ropponmatsu 4-2-1, Fukuoka 810-8560, Japan}
\bigskip
\centerline{ ${}^b$ Laboratoire de Physique Th\'eorique et Hautes
Energies \footnote[1]{\it Laboratoire associ\'e au CNRS.}}
\centerline{ Universit\'e Pierre et Marie Curie, Tour 16 1$^{er}$
		\'etage, 4 place Jussieu}
\centerline{75252 Paris cedex 05-France}
\vskip2cm
\noindent
{\bf Abstract.} 
We study the affine ring of the affine Jacobi variety of a 
hyper-elliptic curve.
The matrix construction of the affine hyper-elliptic Jacobi varieties
due to Mumford is used to calculate the character of the affine ring.
By decomposing the character we make several conjectures on the cohomology
groups of the affine hyper-elliptic Jacobi varieties.
In the integrable system described by the family of
these affine hyper-elliptic Jacobi varieties, the affine ring
is closely related to the algebra of functions on the phase space,
classical observables. 
We show that the affine ring is 
generated by the highest cohomology group over
the action of the invariant vector fields on the
Jacobi variety.

\newpage

\section{ Introduction.}

Our initial motivation is the study of
integrable systems.
Consider an integrable system with
$2n$ degrees of freedom, 
by definition it
possesses $n$ integrals
in involution. The levels of these integrals
are $n$-dimensional tori. This 
is a general description, but the particular
examples of integrable models that we meet in practice 
are much more special. Let us explain how they are organized.

The phase space $\mathcal{M}$ is embedded 
algebraically into the space $\mathbb{R}^N$.
The integrals are algebraic
functions of coordinates in this space. 
This situation allows complexification, the
complexified phase space $\mathcal{M}^{\mathbb{C}}$
is an algebraic affine variety embedded into
$\mathbb{C}^N$. The levels of integrals in the
complexified case allow the following beautiful description.
The systems that we consider are such that with
every one of them one can identify an algebraic
curve $X$ of genus $n$ whose moduli are defined
by the integrals of motion. On the Jacobian $J(X)$ of this curve 
there is a particular divisor $D$ (in this paper we consider the
case when this divisor coincides with the theta divisor, but
more complicated situations are possible).
The level of integrals is isomorphic to the
affine variety $J(X)-D$. The real space $\mathbb{R}^N\subset \mathbb{C}^N$
intersects with every level of integrals
by a compact real sub-torus of $J(X)-D$.

This structure explains 
why the  methods of algebraic geometry are so important
in application to integrable models.
Closest to the present paper account of these
methods is given in the Mumfords's book \cite{mum}.

Let us describe briefly the results of the present paper.
We study the structure of the ring $\mathbf{A}$ of algebraic functions
(observables) on the phase space of certain integrable model.
The curve $X$ in our case is hyper-elliptic.
As is clear from the description given
above this ring of algebraic functions is, roughly, a
product of the functions of integrals of motion by
the affine ring of hyper-elliptic Jacobian.
The commuting vector fields defined by
taking Poisson brackets with the integrals
of motion are acting on $\mathbf{A}$. We shall show that
by the action of these vector-fields the ring $\mathbf{A}$
is generated from finite number of functions corresponding
to the highest nontrivial cohomology group
of the affine Jacobian. We conjecture the
form of the cohomology groups in every degree and demonstrate
the consistence of our conjectures with the structure
of the ring $\mathbf{A}$.

Finally we would like to say
that this relation to cohomology groups became clear 
analyzing the results of papers \cite{bbs} and \cite{toda}
which deal with quantum integrable models. 
Very briefly the reason for that is as follows.
The quantum observables are in one-to-one
correspondence with the classical ones.
Consider a matrix element of some observable
between two eigen-functions of Hamiltonians.
An eigen-functions written in ``coordinate'' representation
(for ``coordinates'' we take the angles on the torus)
must be considered as proportional to square-root
of the volume form on the torus. The matrix element is written as 
integral with respect to ``coordinates'', the product
of two eigen-functions gives a volume form on the
torus, and the  operator itself can be considered, at least semi-classically, 
as 
a multiplier in front of this volume form, i.e.
as coefficient of some differential top form
on the torus which is the same as the form of one-half
of maximal dimension on the phase space. 
Further, those operators which correspond
to ``exact form'' have vanishing matrix elements.
This is how the relation to the cohomologies appears.

The paper, after the introduction, consists of five sections and
six appendices which contain technical details and some proofs.

In section 2 we recall the standard construction
of the Jacobi variety which is valid for any 
Riemann surface.

An algebraic construction of the affine Jacobi variety $J(X)-\Theta$
of a hyper-elliptic curve $X$ is reviewed in section 3 following the book
\cite{mum}vol.II.
This construction is specific to hyper-elliptic curves or more
generally spectral curves \cite{Beauv}.

In section 4 we study the affine ring of $J(X)-\Theta$ using the description in
section 3.
The main ingredient here is the character of the affine ring.
To be precise we consider the ring $\mathbf{A}_0$ corresponding to the
most degenerate curve $y^2=z^{2g+1}$.
The ring $\mathbf{A}$ and the affine ring $\mathbf{A}_f$ of $J(X)-\Theta$
for a non-singular $X$ can be studied using $\mathbf{A}_0$.
It is important that $\mathbf{A}_0$ is a graded ring and the character
$\text{ch}(\mathbf{A}_0)$ is defined.
We calculate it by determining explicitly a $\mathbb{C}$-basis of 
$\mathbf{A}_0$.
The relation between $\mathbf{A}_0$ and $\mathbf{A}_f$ for a non-singular $X$
is given in Appendix E.

A set of commuting vector fields acting on $\mathbf{A}$ is introduced
in section 5.
This action descends to the quotients $\mathbf{A}_0$ and $\mathbf{A}_f$.
The action of the vector fields coincides with the action of invariant
vector fields on $J(X)$.
With the help of these vector fields we define the de Rham type
complexes $(\mathbf{C}^\ast,d)$, $(\mathbf{C}^\ast_0,d)$,
$(\mathbf{C}^\ast_f,d)$ with the coefficients in $\mathbf{A}$,
$\mathbf{A}_0$, $\mathbf{A}_f$ respectively.
The complex $(\mathbf{C}^\ast_f,d)$ is nothing but the algebraic de Rham
complex of $J(X)-\Theta$ whose cohomology groups are known 
to be isomorphic to the singular cohomology groups of $J(X)-\Theta$.
What is interested for us is the cohomology groups
of $(\mathbf{C}^\ast_0,d)$.
We calculate the $q$-Euler characteristic of $(\mathbf{C}^\ast_0,d)$
and show that it coincides with the quotient of $\text{ch}(\mathbf{A}_0)$
by the character $\text{ch}({\cal D})$ of the space of commuting vector fields.
Then, by the Euler-Poincar\'e principle, 
$\text{ch}(\mathbf{A}_0)/\text{ch}({\cal D})$
is found to be expressible as the alternating sum of the characters
of cohomology groups of $(\mathbf{C}^\ast_0,d)$.
Decomposing independently the explicit formula of $\text{ch}(\mathbf{A}_0)$
into the alternating sum, we make conjectures on the cohomology groups of
$(\mathbf{C}^\ast_0,d)$ which are formulated in the next section.

In section six and in Appendix B we study the singular 
homology and cohomology groups of $J(X)-\Theta$.
The Riemann bilinear relation plays an important role here.
We formulate conjectures on the cohomology groups of
$(\mathbf{C}^\ast,d)$, $(\mathbf{C}^\ast_0,d)$,
$(\mathbf{C}^\ast_f,d)$.

\vskip5mm

\noindent
{\bf Acknowledgements.}
This work was begun during the visit of one of the authors (A.N.)
to LPTHE of Universit\'e Paris VI and VII in 1998-1999.
We express our sincere gratitude to this institution
for generous hospitality.
A.N thanks K. Cho for helpful discussions.

\section{Hyper-elliptic curves and their Jacobians.}
Consider the hyper-elliptic curve $\S$ of genus $g$
described by the equation:
\begin{align}
y^2 =f(z),
\non
\end{align}
where
\begin{align}
f(z)=z^{2g+1}+f_1z^{2g}+ \cdots +f_{2g+1}.
\label{i}
\end{align}
The hyper-elliptic
involution $\sigma$ is defined by
\begin{align}
\sigma(z,y)=(z, -y).
\non
\end{align}
The Riemann surface $\S$  can be realized as two-sheeted
covering of the $z$-sphere with the quadratic branch points
which are zeros of the polynomial
$f(z)$ and $\infty$.

A basis of holomorphic
differentials is given by:
$$
\mu _j =z^{g-j}\frac{dz}{y},\qquad j=1,\cdots, g.
$$

Choose a canonical homology basis of $X$:
$\a _1,\cdots ,\a _{g}$ , $ \b _1, \cdots ,\b _{g}$ .
The basis of normalized differentials is defined as
$$
\omega _i= \sum\limits _{j=1}^g(M^{-1})_{ij}\mu _j,
$$
where the matrix $M$ consists of $\a$-periods
of holomorphic differentials $\mu_i$:
\begin{align}
&M_{ij}=\int\limits _{\a _j}\mu _i,
\qquad i,j=1\cdots ,g.
\label{ii}
\end{align}
The period matrix
$$
B _{ij}=\int\limits _{\b _i}\omega _j
$$
defines a point $B$ in the Siegel
upper half space:
$$
B_{ij}=B_{ji},\qquad \text{Im}(B)>0.
$$

The Jacobi variety of $\S$ is a $g$-dimensional
complex torus: 
$$
\J=\frac {\mathbb{C}^{g}}
{\mathbb{Z}^{g} + B\mathbb{Z}^{g}}.
$$
The Riemann theta function associated with $\J$ is defined by
$$
\theta (\zeta)=\sum\limits _{m\in\mathbb{Z}^{g}}\text{exp}\ 2\pi i
\(\textstyle{\frac{1}{2}}\ {}^t mB m +{}^t m \zeta\),
$$
where $\zeta \in \mathbb{C}^{g}$. The theta function satisfies
$$
\theta (\zeta +m+B n)=\text{exp}\ 2\pi i
\(-\textstyle{\frac{1}{2}}\ {}^t nB n -{}^t n \zeta\) \ \theta (\zeta ),
$$
for $m,n\in\mathbb{Z}^{g}$.
 
Consider the symmetric product of $\S$, the quotient of the
product space by the action of the symmetric group:
$$
\S(n)=\S^n/S_n.
$$
The Abel transformation defines the map
$$
\S(g) \stackrel{a}{\longrightarrow} \J
$$
explicitly given by
\begin{align}
&w _j=\sum\limits _{k=1}^{g}\int\limits _{\infty} ^{p_k}\omega _j+\Delta,
\non
\end{align}
where $p_1, \cdots ,p_g$ are points of $\S$,
$\Delta $ is the Riemann characteristic 
corresponding to the choice
of $\infty $ for the
reference point. In the present case $\Delta $ is a
half-period because $\infty$ is a branch point \cite{mum}. 

The divisor $\Theta$ is 
the $(g-1)$-dimensional subvariety of
$\J$ defined by
\begin{align}
\Theta=\{w\ |\ \theta (w) =0\}.
\label{double}
\end{align}
The main subject of our study is the ring $A$ of meromorphic
functions on $\J$ with singularities only on $\X$.
The simple way to describe this ring is provided by theta
functions:
\begin{align}
&A=\bigcup\limits_{k=0}^{\infty}
\Big(\frac{\Theta _{k}(w)}{\theta (w)^{k}}\Big),
\label{filter1}
\end{align}
where $\Theta _{k}$ is the space of theta functions of order $k$ i.e.
the space of regular functions on $\mathbb{C}^g$ satisfying
$$
\theta _{k} (w +m+B n)=\text{exp} \ 2k\pi i
\(-\textstyle{\frac{1}{2}}\ {}^t nB n -{}^t n w\) \ \theta _{k} (w).
$$
There are $k^g$ linearly independent theta functions of
order $k$.

Let us discuss the geometric meaning of the ring $A$.
It is well known that with the help of theta functions
one can embed the complex torus $\J$ into the complex projective
space as a non singular algebraic subvariety. 
It can be done, for example, using theta functions of third order:
\newline
1. $3^g$ theta functions of third order define an embedding of $\J$ into
the complex projective space
$\mathbb{P}^{3^g-1}$,
\newline
2. a set of homogeneous algebraic equations for these theta functions can
be written, which allows to describe this embedding as algebraic one.

Now consider the functions
$$
\frac{\Theta _{3}(w)}{\theta (w)^{3}}.
$$
Obviously, with the help of these functions, we can embed the
non-compact variety $\J-\X$ into the complex affine space
$\mathbb{C} ^{3^g-1} $.  Denote the coordinates in this space by
$x_1,\cdots ,x_{3^g-1} $, the
affine ring of $\J-\X$ is defined as the ring
$$
\mathbb{C}\ \mathbb[x_1,\cdots ,x_{3^g-1}]/(g_\alpha),
$$
where  $(g_{\alpha})$ is the ideal generated by the polynomials
$\{g_\alpha\}$ such that $\{g_\alpha=0\}$
defines the embedding.
It is known that
the affine ring is the characteristic of the non-compact
variety $\J-\X$ independent of a particular embedding of this
variety into affine space.
Obviously the ring $A$ defined above is isomorphic to the affine ring.
We remark that the above argument on the embedding $J(X)-\Theta$
into an affine space is valid if $(J(X),\Theta)$ is replaced by
any principally polarized abelian variety.

Consider $\S(g)$ 
which is mapped to $\J$ by the Abel map $a$.
The Riemann theorem says that 
$$
\theta (w)=0\qquad \text{iff}\qquad
w =\sum\limits _{j=1}^{g-1}\int\limits _{\infty}^{p_j}\omega +\Delta 
$$
which allows to describe $\X$ in terms of the symmetric product. 
One easily argues that the preimage
of $\X$  under the Abel map is described as
\begin{align}
&D:=D_{\infty}\cup D_0,
\non
\end{align}
where
\begin{align}
&D_{\infty}=\{(p_1,\cdots ,p_g)\in \S(g) \vert
\ p_i=\infty 
\text{ for some}\  i\} ,\label{iv}\\
&D_0=\ \{(p_1,\cdots ,p_g)\in \S(g) \vert \  p_i=\sigma(p_j)
\text{ for some}\ i\ne j\ \}.
\non
\end{align}
The Abel map
is not one-to-one, 
and the compact varieties $\J$ and $\S(g)$
are not isomorphic. However,
the affine varieties $\J-\X$ and 
$
\S(g)-
D
$ 
are isomorphic since the Abel map
$$
\S(g)-D
\ \stackrel{a}{\longrightarrow } \J-\X
$$
is an isomorphism. In what follows we shall
study the affine variety $\S(g)-D\simeq J(X)-\Theta$.

\section{Affine model of hyperelliptic Jacobian.}
Consider a traceless $2\times 2$ matrix
\begin{align}
&
m(z)=
\pmatrix 
a(z) &b(z)
\non\\
c(z) &-a(z) 
\endpmatrix,
\non
\end{align}
where the matrix elements are polynomials of the form:
\begin{align}
&a(z)=a_{\frac{3}{2}}z^{g-1}+
a_{\frac{5}{2}}z^{g-2}+\cdots +a_{g+{\frac{1}{2}}},\label{vii}\\
&b(z)=z^{g}+b_1z^{g-1}+\cdots +b_{g},\non\\
&c(z)=z^{g+1}+c_1z^{g}+c_2z^{g-1}+\cdots +c_{g+1}.
\non
\end{align}
Later we shall set $b_0=c_0=1$.
Consider the affine space $\mathbb{C}^{3g+1}$ with coordinates
$\ a_{\frac{3}{2}},\cdots ,a_{\frac{g+1}{2}}\ $, $\ b_1,\cdots ,b_{g}\ $,
$c_1,\cdots,c_{g+1}$.
Fix the determinant of $m(z)$:
\begin{align}
a^2(z)+b(z)c(z)=f(z),
\label{viii}
\end{align}
where the polynomial $f(z)$ is the same as used above (\ref{i}).
Comparing each coefficient of $z^i$ $(i=0,1,\cdots,2g)$ of (\ref{viii}) 
one gets $2g+1$ different equations. 
In fact the equations (\ref{viii}) define g-dimensional
sub-variety of $\mathbb{C}^{3g+1}$. 
This algebraic variety is isomorphic to $\J-\X$ as shown in the book 
\cite{mum}.
We shall briefly recall the proof.

Consider a matrix $m(z)$ satisfying (\ref{viii}).
Take the zeros of $b(z)$:
$$b(z)=\prod\limits _{j=1}^g(z-z_j)$$
and set
$$
y_j=a(z_j).
$$
Obviously $z_j,y_j$ satisfy the equation
$$
y^2_j=f(z_j),
$$
which defines the curve $\S$. So, we have constructed
a point of $\S(g)$ for every $m(z)$
which satisfies the equations (\ref{viii}). 
Conversely, for a point $(p_1,\cdots ,p_g)$ of $\S(g)$,
construct the matrix $m(z)$ as
\begin{align}
&b(z)=\prod\limits _{j=1}^g(z-z_j) ,
\qquad 
a(z)=\sum\limits _{j=1}^g \ y_j\prod\limits _{k\ne j}\(
\frac{z-z_k}{z_j-z_k}\),
\non\\
&c(z)= \frac{-a(z)^2+f(z)}{b(z)},
\non
\end{align}
where $z_j=z(p_j)$ is the $z$-coordinate of $p_j$.
Considering the function $b(z)$ as a function on
$\S(g)$ one finds that it has
singularities when one of $z_j$ equals $\infty$.
The function $a(z)$ is singular at $z_j=\infty $
and also at the points where $z_i=z_j$ but $y_i=- y_j$.
This is exactly the description of the variety 
$
D
$.
The functions $a(z)$ and $c(z)$ do not add new singularities.
Thus we have the embedding of the affine variety $X(g)-D$ into
the affine space:
\begin{align}
&\S(g)-D \hookrightarrow \mathbb{C}^{3g+1}.
\non
\end{align}
Therefore we can profit from the wonderful property of the
hyper-elliptic Jacobian: it allows an affine embedding into
a space of very small dimension equal to $3g+1$ (compare
with $3^g-1$ which we have for any Abelian variety). Actually,
the space $\mathbb{C}^{3g+1}$ occurs foliated with generic leaves
isomorphic to the affine Jacobians.

\section{Properties of affine ring.}

Consider the free polynomial
ring $\mathbf{A}$:
$$
\mathbf{A}=\mathbb{C}\ [a_{\frac{3}{2}},\cdots ,a_{g+{\frac{1}{2}}},
b_1,\cdots ,b_{g}, c_1,\cdots ,c_{g+1}].
$$
On the ring $\mathbf{A}$ one can naturally introduce a
grading. Prescribe the degree $j$ to any of
generators $a_j$, $b_j$, $c_j$  and extend this definition
to all monomials in $\mathbf{A}$ by
$$
\text{deg}(xy)=\text{deg}(x)+\text{deg}(y).
$$
Every monomial of the ring has positive degree (except for
$1$ whose degree equals $0$). Thus, as a linear space,
$\mathbf{A}$ splits into
$$
\mathbf{A}  =\bigoplus\limits
_{2p\in \mathbb{Z}_+}\mathbf{A} ^{(p)},
$$
where $\mathbf{A}^{(j)}$ is the subspace of degree $j$
and $\mathbb{Z}_+=\{0,1,2,\cdots\}$.
Define the character of $\mathbf{A}$ by
$$
\text{ch}(\mathbf{A})=\sum\limits _ {2p\in \mathbb{Z}_+}
\ q^p\ \text{dim}( \mathbf{A} ^{(p)} ).
$$
Since the ring $\mathbf{A}$ is freely generated by
$a_p$, $b_p$, $c_p$ one easily finds
\begin{align}
\text{ch}(\mathbf{A}) =
\frac {\[\frac 1 2\]} 
{\[g+\frac{1}{2}\]!\ [g]!\ [g+1]!},
\label{chB}
\end{align}
where, for $k\in \mathbb{Z}_+$, 
$$
[k]=1-q^k,\quad [k]!=[1]\cdots [k],
\quad\[k+{\textstyle{\frac{1}{2}}}\]!=
\[{\textstyle{\frac{1}{2}}}\]
\[{\textstyle{\frac{3}{2}}}\]
\cdots
\[k+{\textstyle{\frac{1}{2}}}\].
$$
This important formula allows to control the size of the ring
$\mathbf{A}$.

The relation of the ring $\mathbf{A}$ to the affine ring $A$
is obvious. The latter is the quotient of $\mathbf{A}$ by 
the ideal generated by the relations
$-\text{det}(m(z))=f(z)$ where the coefficients
of $f$ are considered fixed constants.

{From} the point of view of integrable models, it is more natural
to see $f_1,\cdots,f_{2g+1}$ as variables
than complex numbers.
If we assign degree $j$ to the variables $f_j$,
all the equations in (\ref{viii}) are homogeneous.
Consider the polynomial ring 
$$
\mathbf{F}=\mathbb{C}\ [f_1,\cdots ,f_{2g+1}].
$$
The ring $\mathbf{F}$ is graded 
and its character is
$$
\text{ch}(\mathbf{F})=\frac{1}{[2g+1]!}.
$$

The ring $\mathbf{F}$ acts on $\mathbf{A}$, that is,
$f(z)$ acts by the multiplication of the left hand side of (\ref{viii}).
Consider the space $\mathbf{A} _0$
which consists of $\mathbf{F}$-equivalence classes:
$$
\mathbf{A}_0=\mathbf{A}\ /\ (
\mathbf{F}^{\times}\mathbf{A}),
\qquad\mathbf{F}^{\times}=\sum\limits_{i=1}^{2g+1}\mathbf{F}f_i.
$$
Since $\mathbf{F}^{\times}\mathbf{A}$ is a homogeneous ideal of 
$\mathbf{A}$, $\mathbf{A}_0$ is a graded vector space:
$$
\mathbf{A}_0  =\bigoplus\limits
_{2p\in \mathbb{Z}_+}\mathbf{A}_0 ^{(p)}.
$$

One can
consider the space $\mathbf{A}_0$
as a subspace of $\mathbf{A}$
taking a set of homogeneous
representatives of the equivalence classes
(being homogeneous they are automatically of smallest possible degree).
Consider any
homogeneous  $x\in \mathbf{A}$.
One can write $x$ as
$$
x =x^{(0)}+\sum_{i=1}^{2g+1} f_ix_i,
$$
where $x^{(0)}\in \mathbf{A}_0$
and $x_i$ is a homogeneous element in $\mathbf{A}$
satisfying $\text{deg}\, x_i=\text{deg}\, x-i$.
Since the degree of $x_i$ is less than the degree of $x$,
repeating the same procedure for $x_i$ one
arrives, by finite number of steps, at
\begin{align}
&x=\sum h_jx^{(0)}_j\label{x=gx},
\end{align}
where $x_j^{(0)}\in \mathbf{A}_0$,
$h_i\in \mathbf{F}$ and the summation is finite.
There is an $\mathbf{F}$-linear map:
$$
\mathbf{F}\otimes_{\mathbb{C}} \mathbf{A}_0
\stackrel{m}{\longrightarrow }\ \mathbf{A},
$$
which corresponds to multiplying the elements of 
$\mathbf{A}_0$ by elements from $\mathbf{F}$
and taking linear combinations. The above reasoning
shows that   
$\text{Im}(m)=\mathbf{A}$.
Hence 
\begin{align}
\text{ch}(\mathbf{A}_0)
\ge\frac{\text{ch}(\mathbf{A})}{\text{ch}(\mathbf{F})}.
\label{ge}
\end{align}
The equality takes place iff
$\text{Ker}(m)=0$.
We shall see that this is indeed the case.
Informally the equality $\text{Ker}(m)=0$ is a manifestation
of the fact that the space $\mathbb{C}^{3g+1}$ is foliated
into $g$-dimensional sub-varieties, the coordinates $f_j$
describe transverse direction.
The pure algebraic proof of this fact is given by the following
proposition.
\vskip 3mm
\noindent
{\bf Proposition 1. }
{\it The set of elements
\begin{eqnarray}
&&
\prod_{j=1}^gu_{\frac{1+j}{2}}^{i_j}
\prod_{k=1}^gu_{\frac{g+1+k}{2}}^{l_k},
\nonumber
\end{eqnarray}
where}
$$
u_p=\left\{ 
\begin{array}{rl}
a_p,&\quad p=\text{half-integer}\\
b_p,&\quad p=\text{integer},
\end{array}
\right.
$$
{\it is a basis of $\mathbf{A}_0$  as a vector space, 
where $i_1,\cdots,i_g$ are non-negative integers and $l_1,\cdots,l_g$ are
$0$ or $1$.}
\vskip3mm

\noindent
The proof of Proposition 1 is given in Appendix A.
Proposition 1 shows that
\begin{align}
&\text{ch}(\mathbf{A}_0)=
\prod_{j=1}^g  
\frac {1}{ \[\frac{1+j}{2}\]}\prod_{k=1}^g  
\(1+q^{\frac{g+1+k}{2}} \)
=\frac{\[\frac{1}{2}\][2g+1]!}{\[g+{\frac{1}{2}}\]!\ [g]!\ \ [g+1]!}
=\frac{\text{ch}(\mathbf{A})}{\text{ch}(\mathbf{F})},
\label{A=AZ}
\end{align}
which means that $\text{Ker}(m)=0$.
We summarize this in the following:
\vskip 3mm
\noindent
{\bf Proposition 2.} {\it 
As an $\mathbf{F}$ module, $\mathbf{A}$ is a free module,
$\mathbf{A}\simeq \mathbf{F}\otimes_{\mathbb{C}} \mathbf{A}_0$.
In other words
every element $x\in \mathbf{A}$
can be uniquely presented as
a finite sum:
\begin{align}
&x=\sum h_jx^{(0)}_j,
\non
\end{align}
where $\{x^{(0)}_j\}$ is a basis of the 
$\mathbb{C}$-vector space $\mathbf{A}_0$ and $h_j\in \mathbf{F}$.}

\section{Poisson structure and cohomology groups.}

The affine model of hyper-elliptic Jacobian
is interesting for its application
to integrable models. The ring $\mathbf{A}$ that we
introduced in the previous section can be supplied with
Poisson structure. This fact is also important because
introducing the Poisson structure is the first
step towards the quantization.
The Poisson structure 
in question is described in r-matrix
formalism as follows:
\begin{align}
&\{m(z_1)\otimes I,I\otimes m(z_2)\}=[r(z_1,z_2), m(z_1)\otimes I]-
[r(z_2,z_1), I\otimes m(z_2)].
\label{ix}
\end{align}
The r-matrix acting in $\mathbb{C}^2\otimes\mathbb{C}^2$ is
$$
r(z_1,z_2)=\frac{z_2}{z_1-z_2} \(\textstyle{\frac{1}{2}}
\sigma ^3 \otimes \sigma^3+
\sigma ^+ \otimes \sigma^-
+\sigma ^- \otimes \sigma^+\)
+z_2\ \sigma ^-\otimes\sigma ^-,
$$
where $\sigma ^3$, $\sigma ^{\pm}$ are Pauli matrices.

The variables $z_1,\cdots ,z_g$ (zeros of $b(z)$) and $y_j=a(z_j)$
have dynamical meaning of separated variables
\cite{skl}. The Poisson
brackets (\ref{ix}) imply the following Poisson brackets for
the separated variables:
$$
\{z_i,y_j\}=\delta _{i,j} \ z_i.
$$

The determinant $f(z)$ of the matrix $m(z)$ generates 
Poisson commutative
subalgebra:
$$
\{f(z_1),f(z_2)\}=0.
$$
It can be shown that the coefficients
$f_1,f_2,\cdots ,f_g$ and $f_{2g+1}$
belongs to the center of Poisson algebra. 
The Poisson commutative coefficients
$f_{g+1},\cdots ,f_ {2g}$ are the integrals of motion.
Introduce the commuting vector-fields
$$
D _ i h=\{f_{g+i},h\},\qquad i=1,\cdots g.
$$
For completeness let us describe explicitely the
action of these vector-fields on $m(z)$.
Define 
$$
D(z)=\sum _{j=1}^gz^{j-1}D_{g+1-j}.
$$
Then the Poisson brackets (\ref{ix}) imply:
$$
D(z_1)m(z_2)=\frac {1}{z_1-z_2}[m(z_1),m(z_2)]-
[\sigma ^-m(z_1)\sigma ^-,m(z_2)].
$$

One can think of these commuting vector-fields as $D _j=
\frac{\partial}{\partial \tau _j} $
where $\tau _j$ are "times" corresponding
to the integrals of motion $f_{g+j}$.
The "times" $\tau _j$ are coordinates on the Jacobi variety,
they are related to $w$ as follows
$$
\tau =\textstyle{\frac{1}{2}}Mw,
$$
where $M$ is the matrix defined in (\ref{ii}).
We remark that $D_i$ here coincides with $-2D_i$ 
in the Mumford's book \cite{mum}vol.II.
Earlier we have introduced
a gradation on the ring $\mathbf{A}$. We can prescribe the degrees
to the vector-fields $D _ j$ as 
$\text{deg}\(D _ j  \)=j-\frac{1}{2}$
because it can be shown that:
$$ 
D _ j \mathbf{A}^{(p)}\subset \mathbf{A}^{(p+j-\frac{1}{2})}.
$$

Consider the differential forms
\begin{align}
&f_{i_1\ \cdots \ i_k}d\tau _{i_1}
\ \wedge\cdots\wedge d\tau _{i_k}\label{x},
\end{align}
with $ f_{i_1,\cdots ,i_k}\in\mathbf{A}$.
These forms span the linear spaces $\mathbf{C}^k$ for $k=0,\cdots ,g$.
The differential
$$
d=\sum _{j=1}^g d\tau _j D _ j ,
$$
acts from $\mathbf{C}^k$  to $\mathbf{C}^{k+1}$ .
As usual applying $d$ we first apply the vector fields  $D _ j $
to the coefficients of the differential form and then take
exterior product with  $d\tau _j$. 
We have the complex
$$
0\lar
{\mathbf{C}^{0}}
\stackrel{d}{\longrightarrow }
\ {\mathbf{C}^{1}}
\stackrel{d}{\longrightarrow }
\ \cdots
\stackrel{d}{\longrightarrow }
\ {\mathbf{C}^{g-1}}
\stackrel{d}{\longrightarrow }
\ {\mathbf{C}^{g}}
\stackrel{d}{\longrightarrow }\ 0.
$$
The $k$-th cohomology group of this complex is denoted by 
$H^k(\mathbf{C}^\ast)$.
Consider the problem of grading of the spaces $\mathbf{C}^j$.
Clearly we have to prescribe the degree to $d\tau _j$ as
$$
\text{deg}(d\tau _j)=-j+\textstyle{\frac{1}{2}}
$$
in order that $d$ has degree zero.

Consider the spaces 
${\mathbf{C}^k _0}$ spaned by (\ref{x}) with $f_{i_1\ \cdots \ i_k}\in 
\mathbf{A}_0$. 
Since the elements of $\mathbf{F}$
are ``constants'' (commute with $D _i$),
$D_i$ acts on $\mathbf{A}_0$.
So, we have the complex $\mathbf{C}^*_0$:
$$
0\lar
{\mathbf{C}^{0}_0}
\stackrel{d}{\longrightarrow }
\ {\mathbf{C}^{1}_0}
\stackrel{d}{\longrightarrow }
\ \cdots
\stackrel{d}{\longrightarrow }
\ {\mathbf{C}^{g-1}_0}
\stackrel{d}{\longrightarrow }
\ {\mathbf{C}^{g}_0}
\stackrel{d}{\longrightarrow }\ 0.
$$
This complex is graded.
One easily calculates that
\begin{align}
&\text{ch}({\mathbf{C}^{g-j}_0} )=
q^{\frac{1}{2}(j^2-g^2)}
{\textstyle\left[\ \begin{matrix}g\cr j\end{matrix}
\ \right]}\ \text{ch}\({\mathbf{A}_0}\),
\label{char}
\end{align}
where the q-binomial coefficient is defined
as
$$
\left[\ \begin{matrix}g\cr j\end{matrix}\ \right]=
\frac{[g]!}{[j]!\ [g-j]!}.
$$
The differential  $d$ respects the grading. In this case the
$q$-Euler characteristic can be introduced
\begin{align}
&\chi _q\({\mathbf{C}^{*}_0}\)
=\text{ch}\({\mathbf{C}^{0}_0}\)-
\text{ch}\({\mathbf{C}^{1}_0}\)+\cdots
+(-1)^g\text{ch}\({\mathbf{C}^{g}_0}\),
\label{defEu}
\end{align}
which possesses all the essential properties of the usual
Euler characteristic.
Using the formula (\ref{char}) one finds
\begin{align}
\chi _q\({\mathbf{C}^{*}_0}\)
&=
(-1)^g q^{-\frac{1}{2}g^2}\ \[{\textstyle{\frac{2g-1}{2}}}\]!
\ \text{ch}\({\mathbf{A}_0}\)
\label{Euler}
\\
&=
(-1)^g q^{-\frac{1}{2}g^2}\ 
\frac{[2g+1]!\ [\frac{1}{2}]}{[g+\frac{1}{2}]\ [g]!\ [g+1]!}.
\non
\end{align}

Consider the cohomology groups $H^k(\mathbf{C}^{*}_0)$.
The vector spaces $H^k(\mathbf{C}^{*}_0)$
inherit a grading from $\mathbf{C}^{j}_0$.
Then
\begin{align}
&\chi _q(\mathbf{C}_0^*)=\text{ch}
(H^0(\mathbf{C}^{*}_0))
-\text{ch}(H^1(\mathbf{C}^{*}_0))+\cdots
+(-1)^g\text{ch}(H^g(\mathbf{C}^{*}_0)).
\non 
\end{align}

The $q$-number in (\ref{Euler}) has finite limit for $q\to 1$:
\begin{align}
\lim_{q\rightarrow 1}\chi_q (\mathbf{C}^{*}_0)
=
(-1)^g\frac{(2g)!}{(g)!\ (g+1)!}=(-1)^g\(
{\textstyle\binom{2g}{g}-\binom{2g}{g-1}}\).
\label{Eu0}
\end{align}
Certainly the fact that the $q$-Euler characteristic
has a finite limit does not mean 
that cohomology groups are finite-dimensional,
but we believe that this is the case.
So, we put forward
\vskip 3mm
\noindent
{\bf Conjecture 1.} {\it The spaces }
$H^k(\mathbf{C}^{*}_0)$ {\it are finite-dimensional}.
\vskip 3mm
\noindent
More explicitly the cohomology groups will be discussed in the next
section.

In the situation under consideration there is an important
connection between the algebra $\mathbf{A}$  and the
highest cohomology group $H^g(\mathbf{C}^{*}_0)$.
\vskip 3mm
\noindent
{\bf Proposition 3.} {\it Consider
some homogeneous representatives of a basis of the space 
$H^{g}(\mathbf{C}^{*}_0)$:
$$
h_\alpha \ d\tau_1\wedge\cdots \wedge d\tau _g.
$$
Arbitrary $x\in {\mathbf{A}_0} $ can be presented in the form}
\begin{align}
x=
\sum\limits _{\alpha }P_{\alpha }( D_1,\cdots D_g)
h_{\alpha },
\label{f=dh}
\end{align}
{\it where 
$P_{\alpha }(D_1,\cdots D_g)$ are 
polynomials in
$D_1,\cdots,D_g $ 
with $\mathbb{C}$-number coefficients.}
\vskip 3mm
\noindent
{\it Proof.}
For $x\in {\mathbf{A}_0} $ construct
$$
\Omega =x\ d\tau_1\wedge\cdots\wedge d\tau _g
\in {\mathbf{C}^g_0}.
$$
By the definition of cohomology group we have
$$
\Omega =\Omega _0+d\Omega ',\qquad \Omega _0\in 
H^{g}(\mathbf{C}^{*}_0),\qquad \Omega '
\in {\mathbf{C}^{g-1}_0},
$$
which implies that
$$
x= h+\sum D_i x_i,
$$
with $h$ such that 
$\Omega _0=h\ d\tau_1\wedge\cdots \wedge d\tau _g$, 
$x_i\in {\mathbf{A}_0}$.
Apply the same procedure to $x_i$ and go on along the same lines.
The resulting representation (\ref{f=dh}) will be
achieved in finite number of steps for the reason of grading.
QED.
\vskip3mm

Let us introduce the notation
$${\cal D}=\mathbb{C}\ [D _1,\cdots,D _g].$$
We shall call the expressions of the type (\ref{f=dh}) the
${\cal D}$-descendents of $\{h_{\alpha }\}$.
The interesting question concerning the formula (\ref{f=dh})
is whether such representation is unique for any $x$.
The answer is that it is not the case, and to understand
why it is so we have to return to the formula  (\ref{Euler})
which can be rewritten as follows:
\begin{align}
&q^{-\frac{1}{2}g^2}
\text{ch}\({\mathbf{A}_0}\)=
\frac{1}{\[g-\frac{1}{2}\]!}
\text{ch}(H^{g}(\mathbf{C}^{*}_0))-\non\\&
-
\frac{1}{\[g-\frac{1}{2}\]!}
\text{ch}(H^{g-1}(\mathbf{C}^{*}_0))
+
\frac{1}{\[g-\frac{1}{2}\]!}
\text{ch}(H^{g-2}(\mathbf{C}^{*}_0))-\cdots.
\label{fff}
\end{align}
Obviously, the first term in the RHS represents the character of the
space of all 
${\cal D}$-descendents of $\{h_{\alpha }\}$ 
(recall that  the degree of $D_j$ equals $j-\frac{1}{2}$). 
This is equivalent to saying that
the first term has the same character as the space 
generated freely over ${\cal D}$
by $H^{g}(\mathbf{C}^{*}_0)$:
$$
\frac{1}{\[g-\frac{1}{2}\]!}
\text{ch}(H^{g}(\mathbf{C}^{*}_0))
=\text{ch}({\cal D})\text{ch}(H^{g}(\mathbf{C}^{*}_0))
=\text{ch}
\big({\cal D}\otimes_{\mathbf C}H^{g}(\mathbf{C}^{*}_0)\big).
$$

The existence of the second term of the RHS of (\ref{fff}) 
implies that, in $\mathbf{A}_0$, there are linear relations among 
${\cal D}$-descendents of $\{h_{\alpha }\}$ and
they are parametrized by the second term.
The third term explains that there are relations among linear relations 
counted by the second term of the RHS of (\ref{fff}) and so on.
This is nothing but the usual argument of constructing
a resolution of a module. In the present case it is actually possible to
construct a free resolution of $\mathbf{A}_0$ as a ${\cal D}$ module
assuming some conjectures.
The construction of the free resolution is given in Appendix F.


Combining Proposition 2 and Proposition 3 one arrives at
\vskip 3mm
\noindent
{\bf Proposition 4.}
{\it Let $\{h_\alpha\}$ be the same as in Proposition 3.
Then every $x\in\mathbf{A}$ can be presented as
\begin{align}
&x=
\sum\limits _{\alpha }
\mathbf{P}_{\alpha }\(D _1,\cdots,D_ g\)
h_{\alpha },
\label{f=dh1}
\end{align}
where 
$\mathbf{P}_{\alpha }\(D _1,\cdots,D _ g\)$ are 
polynomials in $D_1$,...,$D_g$ with  coefficients from $\mathbf{F}$.}
\vskip 3mm
\noindent
{\it Proof.} We shall prove the proposition by the induction
on the degree of $x$. Since $\mathbf{A}^{(p)}=\{0\}$ for $p<0$, the beginning
of induction obviously holds. Suppose that the proposition is true 
for all elements of degree less than $\text{deg}(x)$.
By Proposition 2, there exist $x_j$ such that
$$
x=x_0+\sum_{i=1}^{2g+1}f_ix_i,
\quad
x_j\in\mathbf{A}_0,
$$
where $\text{deg}(x)=\text{deg}(x_0)$ and 
$\text{deg}(x_i)=\text{deg}(x)-i<\text{deg}(x)$ for $i>0$.
By Proposition 3, there exist polynomials $P_\alpha(D_1,\cdots ,D_g)$
with the coefficients in $\mathbb{C}$  such that
$$
x_0=\sum P_\alpha(D_1,\cdots ,D_g)h_\alpha+\sum_{i=1}^{2g+1}f_iy_i,
\quad
y_i\in \mathbf{A},
$$
where $\text{deg}\, y_i=\text{deg}\, x_0-i$.
Since $\text{deg}(y_i)<\text{deg}(x_0)$, $x$ can be written in the form
(\ref{f=dh1}) by the induction hypothesis. 
QED.
\vskip3mm

Proposition 4  represents the most important result
of this paper. The possibility of presenting every algebraic
function on the phase space of the integrable model in the
form (\ref{f=dh1}) starting from finite number of functions
$\{h_{\alpha}\}$, which are representatives of the highest
cohomology group, is important both in classical and in quantum case.
The description of null-vectors follows from the one 
given above because $D_i$ commute with $f_i$.

\section{Conjectures on cohomology groups.}

In the previous section we have seen that the cohomology group
$H^g(\mathbf{C}^{*}_0)$ is important for describing
the algebra $\mathbf{A}$. This cohomology group is rather exotic,
since the complex $\mathbf{C}^\ast_0$ corresponds to the case
when the algebraic curve $X$ is singular, that is, $y^2=z^{2g+1}$.
In this section we first discuss the relation between 
$H^k(\mathbf{C}^{*}_0)$ and the singular cohomology groups
of the non-singular affine Jacobi variety $J(X)-\Theta$.
For a set of complex numbers $f^0=(f_1^0,\cdots,f_{2g+1}^0)$ we set
\begin{eqnarray}
&&
\mathbf{A}_{f^0}=\mathbf{A}\otimes_{\mathbf{F}}\mathbb{C}_{f^0},
\quad
\mathbb{C}_{f^0}=\mathbf{F}/\sum_{i=1}^{2g+1}\mathbf{F}(f_i-f_i^0),
\nonumber
\end{eqnarray}
and $f^0(z)=z^{2g+1}+f^0_1z^{2g}+\cdots+f^0_{2g+1}$.
In the case when all $f^0_i=0$, $\mathbf{A}_{f^0}=\mathbf{A}_0$.
If the curve $X$: $y^2=f^0(z)$ is non-singular, $\mathbf{A}_{f^0}$ is
isomorphic to the affine ring of $J(X)-\Theta$.
Since $d$ commutes with $\mathbf{F}$, the complex $(\mathbf{C}^\ast,d)$
induces the complex $(\mathbf{C}^\ast_{f^0},d)$, where
$$
\mathbf{C}^k_{f^0}=
\mathbf{C}^k\otimes_{\mathbf{F}}\mathbb{C}_{f^0}
=\sum_{i_1<\cdots<i_k}\mathbf{A}_{f^0}d\tau_{i_1}
\wedge\cdots\wedge d\tau_{i_k}.
$$
Recall that
\begin{eqnarray}
&&
H^g(\mathbf{C}^\ast)=\frac{\mathbf{C}^g}{d\mathbf{C}^{g-1}},
\quad
H^g(\mathbf{C}^\ast_{f^0})=\frac{\mathbf{C}^g_{f^0}}{d\mathbf{C}^{g-1}_{f^0}}.
\nonumber
\end{eqnarray}
Thus, tensoring $\mathbb{C}_{f^0}$ to the exact sequence
$$
\mathbf{C}^{g-1}\stackrel{d}{\lar}\mathbf{C}^g\lar H^g(\mathbf{C}^\ast)\lar0,
$$
we have
$$
H^g(\mathbf{C}^\ast)\otimes_{\mathbf{F}}\mathbb{C}_{f^0}
\simeq
H^g(\mathbf{C}^\ast_{f^0}),
$$
for any $f^0$.
By Proposition 4 we have
$$
H^g(\mathbf{C}^\ast)=\sum_\alpha\mathbf{F}\Omega _\alpha,
$$
where $\{\Omega _\alpha\}$ are representatives of $H^g(\mathbf{C}^\ast_0)$ in
$\mathbf{C}^g$. In other words there is a surjective map of 
$\mathbf{F}$-modules:
$$
\mathbf{F}\otimes_{\mathbb{C}}H^g(\mathbf{C}^\ast_0)
\lar
H^g(\mathbf{C}^\ast).
$$
We conjecture that this map is in fact injective.
In general we put forward the following conjecture.
\vskip3mm

\noindent
{\bf Conjecture 2.} (1) {\it $H^k(\mathbf{C}^\ast)$ 
is a free $\mathbf{F}$-module for any $k$.}

\hskip 2cm
(2) {\it $H^k(\mathbf{C}^\ast)\otimes_{\mathbf{F}}\mathbb{C}_{f^0}
     \simeq H^k(\mathbf{C}^\ast_{f^0})$ for any $k$ and $f^0$.}
\vskip5mm

\noindent
Notice that Conjecture 2 implies, in particular, that
$$
H^k(\mathbf{C}^\ast)
\simeq 
\mathbf{F} \otimes_{\mathbb{C}} H^k(\mathbf{C}^\ast_{0}).
$$
It is known, by the algebraic de Rham theorem ({\it cf.} \cite{GH}), that,
if $X$ is non-singular,
$$
H^k(\mathbf{C}^\ast_{f^0})
\simeq
H^k(J(X)-\Theta,\mathbb{C}),
$$
where the RHS is the singular cohomology group of $J(X)-\Theta$.
Thus we have
\vskip3mm

\noindent
{\bf Corollary of Conjecture 2.}{\it There
is an isomorphism:} 
$$
H^k(\mathbf{C}^{*}_0)\simeq H^k(\mathbf{C}^{*}_{f^0}),
$$
{\it for any $f^0$. 
In particular, for any non-singular hyper-elliptic curve} $X$,
$$
H^k(\mathbf{C}^{*}_0)\simeq H^k(X(g)-D, \mathbb{C}).
$$
\vskip 3mm

Notice that Conjecture 1 follows from Conjecture 2 because
the singular cohomology groups of a non-singular affine variety 
are finite-dimensional.
We shall comment more on Conjecture 2 later, for the moment
let us concentrate on the singular cohomology groups 
of $X(g)-D$ for a non-singular $X$.

Consider the affine curve
$\S_{\text{aff}}=\S-\{\infty\}$ and its symmetric
powers: 
$$
\S_{\text{aff}}(n)=\S_{\text{aff}}^n/S_n.
$$
Since $\S_{\text{aff}}$ is affine and connected,
$$
H^p(\S_{\text{aff}},\mathbb{C})=0, \quad p\geq 2,
\quad 
\text{dim}\ H^0(\S_{\text{aff}},\mathbb{C})=1.
$$
The cohomology group $H^1(\S_{\text{aff}},\mathbb{C})$
is $2g$ dimensional and it is generated by 
$$
\mu _j =z^{g-j}\frac{dz}{y},\qquad j=-g+1,\cdots, g,
$$
in the algebraic de Rham cohomology description of 
$H^1(\S_{\text{aff}},\mathbb{C})$.
On $H^1(\S_{\text{aff}},\mathbb{C})$ there is a skew-symmetric
bilinear form:
$$
\lambda _1\circ \lambda _2 =\text{res}_{p=\infty}\(\lambda _1(p)
\int ^p\lambda _2\).
$$
Canonical basis 
$\nu _j$, $j=-g+1,\cdots ,\nu _g$, with respect to this form,  
is defined as one satisfying
$$
\nu _i\circ\nu _j=\frac{4}{j-i}\  \delta _{i+j,1}.
$$
A particular example of such basis is given in Appendix B.

As in the case of the compact curve $\S$, the cohomology groups 
of the symmetric products $\S_{\text{aff}}(n)$ is described as 
the $S_n$ invariants,
$
H^\ast(\S_{\text{aff}}(n),\mathbb{C})
\simeq 
H^\ast(\S_{\text{aff}}^n,\mathbb{C})^{S_n}
$ 
({\it cf.} (1.2) in \cite{Mac}).
If we define 
\begin{align}
&
\tilde{\mu}_i=\mu _i^{(1)}+\cdots+\mu _i^{(n)},
\label{mut}
\\
&
\mu _i^{(k)}=1\otimes\cdots\otimes \stackrel{k}{\breve{\mu _i}} \otimes\cdots\otimes 1\in
H^\ast(X_{\text{aff}},\mathbb{C})^{\otimes n},
\non
\end{align}
then they generate the cohomology ring 
$H^\ast(\S_{\text{aff}}(n),\mathbb{C})$.
Obviously
\begin{align}
&
H^1(\S_{\text{aff}}(n),\mathbb{C})
\simeq 
\mathbb{C}\tilde{\mu}_{-g+1}
\oplus\cdots\oplus
\mathbb{C}\tilde{\mu}_{g},
\non
\\
&
H^k(\S_{\text{aff}}(n),\mathbb{C})
\simeq 
\bigwedge^k H^1(\S_{\text{aff}}(n),\mathbb{C}).
\label{wedge}
\end{align}


Recall that $D=D_0\cup D_{\infty}$. Obviously, $X_{\text{aff}}(g)=
X(g)-D_{\infty}$. Hence there is a map from 
$H^k(X_{\text{aff}}(g),\mathbb{C})$
to $H^k(X(g)-D,\mathbb{C})$. In the Appendix B we prove the following:
\vskip 3mm
\noindent
{\bf Proposition 5.} {\it 
Consider the natural map:}
$$
H^k(X_{\text{aff}}(g),\mathbb{C})\ \stackrel{i'}{\longrightarrow }
{H}^k(X(g)-D,\mathbb{C}).
$$
{\it The kernel of this map is described as follows:}
$$
\text{Ker}(i')=\omega\wedge H^{k-2}(X_{\text{aff}}(g),\mathbb{C}),
$$
{\it where
\begin{eqnarray}
&&
\omega =\textstyle{\frac{1}{4}}\sum\limits _{k=1}^g (2k-1)
\ \tilde{\nu} _k\wedge \tilde{\nu} _{-k+1}.
\label{omega}
\end{eqnarray}}
\vskip5mm
\noindent
We remark that $\omega$ does not depend on the choice of the canonical
basis $\{\nu_j\}$.
By Proposition 5 the map $i'$ induces an injective map:
\begin{align}
W^k:=
H^k (X_{\text{aff}}(g),\mathbb{C})/
\big(
\omega
\wedge  
H^{k-2}(X_{\text{aff}}(g),\mathbb{C})\big)
\hookrightarrow
{H}^k(X(g)-D,\mathbb{C}).
\label{W}
\end{align}
We can make $W^k$ a graded vector space by prescribing the
degrees to differential forms as
$$
\text{deg}(\tilde{\mu} _j)=-j+{\textstyle{\frac{1}{2}}},
\qquad \text{deg}(\omega )=0.
$$
{From} (\ref{wedge}) one easily finds:
\begin{align}
&\text{ch}(W^k)=R_k-R_{k-2},\non\\
&R_k=q^{\frac{1}{2}k(k-2g)}
{\textstyle\left[\begin{matrix}2g\cr k\end{matrix}\right]}.
\non
\end{align}

Now we put forward the strong conjecture that the map (\ref{W}) is 
in fact surjective and the character of $W^k$ defined here coincides
with the character of $H^k(\mathbf{C}_0^\ast)$:
\vskip 3mm
\noindent
{\bf Conjecture 3.} (1) $W^k \simeq H^k(X(g)-D,\mathbb{C})$ for $0\leq k\leq g$.

\hskip 2cm
(2) $\text{ch}(W^k)=\text{ch}(H^k(\mathbf{C}_0^\ast))$.
\vskip 3mm
\noindent
What does it mean? The divisor $D$ consists of $D_{\infty}$ and
$D_0$. The forms from $W^k$ describe the part of cohomology groups
with singularities on $D_{\infty}$ only. Our conjecture is
that this part exhausts the whole space of the cohomology groups, i.e.
that adding exact forms one can move singularities of any form from
$D_0$ to $D_{\infty}$. This is a strong
statement which is rather difficult to prove. The first non-trivial
case is $g=3$ for which we were able to prove Conjecture 3.
The details of it will be published elsewhere.

For $k=1$ we can prove Conjecture 3 (1) for any $g$.
The proof is given in Appendix D.

Let us now present a simple calculation which shows that Conjectures 2, 3
are consistent with the calculation of $q$-Euler characteristic of 
$\mathbf{C}^*_0$. 
Indeed
\begin{align}
&
\text{ch}(W ^0)-\text{ch}(W ^1)+\text{ch}(W ^2)-\cdots+
(-1)^g\text{ch}(W ^g)
\non\\
=&
(R_0)-(R_1)+(R_2-R_0)-(R_3-R_1)+\cdots
+(-1)^g(R_g-R_{g-2})
\non\\
=&
(-1)^g(R_g-R_{g-1})=\chi _q(\mathbf{C}^*_0).
\non
\end{align}
This calculation was actually the starting point for Conjectures 2, 3.
Certainly it does not prove anything, but it shows remarkable
consistence between different calculations performed in this paper.

In order to understand the magic of the hyper-elliptic case
it is instructive to compare it with the case of an Abelian
variety in generic when the divisor $\Theta$ is non-singular
(which rarely the case for Jacobians of algebraic
curves). 
As explained in the Appendix C
in the latter case the following can be proven:
\begin{align}
&
W^k\simeq H^{k}(J-\Theta,\mathbb{C}),
\quad 
k\leq g-1,
\nonumber
\\
&
W^g \hookrightarrow H^{g}(J-\Theta,\mathbb{C}).
\nonumber
\end{align}
Actually for $g\ge 3$ the space $H^{g}(J-\Theta,,\mathbb{C})$
is bigger than $W^g$, the difference of dimensions being
$$
\text{dim}\ H^{g}(J-\Theta,\mathbb{C}) -\text{dim}\ W^g
=
g!-\frac{(2g)!}{g!(g+1)!}.
$$
We would conjecture that the equality 
$H^{g}(J-\Theta,\mathbb{C})\simeq W^g$
specifies hyper-elliptic Jacobians.

\section{Appendix A. Proof of Proposition 1}

We have to determine the basis of $\mathbf{A}_0$.
Recall that
$$
\mathbf{A}_0\simeq\mathbf{A}/\sum_{j=1}^{2g+1}f_j\mathbf{A}.
$$
Write the equations $f_1=\cdots =f_{2g+1}=0$ explicitly:
\begin{eqnarray}
&&
c_k+\sum_{i+j=k, j\neq k}b_ic_j+
\sum_{i+j=k-1}a_{i+\frac{1}{2}}a_{j+\frac{1}{2}}=0,
\quad
1\leq k\leq g+1,
\label{defeq1}
\\
&&
\sum_{i+j=k}b_ic_j+
\sum_{i+j=k-1}a_{i+\frac{1}{2}}a_{j+\frac{1}{2}}=0,
\quad
g+2\leq k\leq 2g+1.
\label{defeq2}
\end{eqnarray}
{From} (\ref{defeq1}) $c_k$ $(1\leq k\leq g+1)$ can be solved
by $b_i$, $a_j$.
It is sometimes convenient to use the following notation;
\begin{eqnarray}
&&
u_{\frac{g}{2}+1}^{m_1}\cdots u_{g+\frac{1}{2}}^{m_g}
=[-m_1,\cdots,-m_g],
\nonumber
\\
&&
B_0=\mathbb{C}\ [u_1,u_{\frac{3}{2}},\cdots,u_{\frac{g+1}{2}}].
\nonumber
\end{eqnarray}
We use both this bracket notation and the $u_j$ notation
to denote a monomial.
The ring $B_0$ is considered as a coefficient in the sequel.

Let us first prove
\vskip 3mm
\noindent
{\bf Proposition A. }
{\it In the ring $A_0$ any monomial $[-m_1,\cdots,-m_g]$ can be written as
a linear combination of monomials of the form $[-n_1,\cdots,-n_g]$,
$n_1,\cdots,n_g=0,1$ with the coefficient in $B_0$.}
\vskip3mm
\noindent
{\it Proof.} Define the degree of $[-m_1,\cdots,-m_g]$ as that of the
monomial:
\begin{eqnarray}
&&
\hbox{deg} [-m_1,\cdots,-m_g]=
\sum_{k=1}^g m_k\frac{g+1+k}{2}.
\nonumber
\end{eqnarray}
We define a total order on the set of monomials $\{[-m_1,\cdots,-m_g]\}$
by the following rule.
Let $P=[-m_1,\cdots,-m_g]$, $P^\prime=[-m_1^\prime,\cdots,-m_g^\prime]$.
\begin{description}
\item[1.] If $\hbox{deg}(P)<\hbox{deg}(P^\prime)$, then $P<P^\prime$.
\item[2.] If $\hbox{deg}(P)=\hbox{deg}(P^\prime)$, compare $P$ and $P^\prime$
by the lexicographical order from the left.
\end{description}
Notice that the product of two elements  
$P=[-m_1,\cdots,-m_g]$
and $P^\prime=[-m_1^\prime,\cdots,-m_g^\prime]$ is expressed as
\begin{eqnarray}
&&
PP^\prime=[-(m_1+m_1^\prime),\cdots,-(m_g+m_g^\prime)].
\nonumber
\end{eqnarray}
The following property obviously holds.
\vskip 3mm
\noindent
{\bf Lemma A 1.}
{\it For monomials $P_1,P_2,P_3$, if $P_1<P_2$ then $P_1P_3<P_2P_3$.}
\vskip 3mm
\noindent
{From} (\ref{defeq1}) 
\begin{eqnarray}
&&
c_k=-u_k+\cdots,
\quad
1\leq k\leq g+1,
\nonumber
\end{eqnarray}
where $\cdots$ part does not contain $u_k$.
Then from (\ref{defeq2}) 
\begin{eqnarray}
&&
u_{\frac{k}{2}}^2=\cdots,
\quad
g+2\leq k \leq 2g+1.
\label{square}
\end{eqnarray}
The next lemma describes what kind of monomials appear in the 
right hand side of (\ref{square}).
\vskip 3mm
\noindent
{\bf Lemma A2. }
{\it The right hand side of (\ref{square}) is a linear combination of
elements of the form $u_l$, $u_lu_m$, $u_lu_mu_n$ with the coefficients
in $B_0$.}
\vskip 3mm
\noindent
{\it Proof.} It is sufficient to show that the 
term like $xu_{i_1}\cdots u_{i_r}$, $r\geq 4$, $x\in B_0$
does not appear in the expression.
Since (\ref{defeq1}), (\ref{defeq2}) are homogeneous,
if $x\neq 0$ and homogeneous, then
\begin{eqnarray}
&&
\hbox{deg}( xu_{i_1}\cdots u_{i_r})=k,
\nonumber
\end{eqnarray}
where we take into account the degree of $x$.
Since $i_1,\cdots,i_r\geq g/2+1$ and $2g+1\geq k$,
\begin{eqnarray}
&&
\hbox{deg} (xu_{i_1}\cdots u_{i_r})
\geq
\hbox{deg} (u_{i_1}\cdots u_{i_r})
=i_1+\cdots+i_r
\geq
4\(\frac{g}{2}+1\)
>k.
\nonumber
\end{eqnarray}
Thus $x=0$.
QED.
\vskip 3mm
\noindent
{\bf Lemma A3. }
{\it In each of the cases in Lemma A2 we have the 
following statements, where $x$ is a homogeneous element in $B_0$.}
\begin{description}
\item[1.] If $u_{k/2}^2=xu_l+\cdots$, then $u_{k/2}^2>u_l$.
\item[2.] If $u_{k/2}^2=xu_lu_m+\cdots$, then $u_{k/2}^2>u_lu_m$.
\item[3.] If $u_{k/2}^2=xu_lu_mu_n+\cdots$, then $u_{k/2}^2>u_lu_mu_n$.
\end{description}
\vskip3mm
\noindent
{\it Proof.} 1. Since $\hbox{deg}(u_{k/2}^2)
=k\geq g+2>g+1/2\geq l=\hbox{deg}(u_l)$,the claim follows.

2. If $k>l+m$, there is nothing to be proved.
Suppose that $k=l+m$. Then $l<k<m$. Thus comparing by the lexicographical
order we have $u_{k/2}^2>u_lu_m$.
The statement of 3 is similarly proved.
QED.
\vskip3mm

Starting from any element $P=[-m_1,\cdots,-m_g]$ we shall show that
$P$ can be reduced to the desired form.
If some $m_j\geq 2$, then rewrite it using $(\ref{square})$.
By Lemma A3 every term in the resulting expression 
is less than $P$.
Repeating this procedure we finally arrive at the linear
combinations of $[-n_1,\cdots,-n_g]$, $n_1,\cdots,n_g=0,1$
with the coefficients in $B_0$.
Thus Proposition A is proved.
QED.
\vskip5mm

By Proposition A we have
\begin{align}
&\text{ch}(\mathbf{A}_0)\leq
\prod_{j=1}^g  
\frac {1}{ \[\frac{1+j}{2}\]}\prod_{k=1}^g  
\(1+q^{\frac{g+1+k}{2}} \)
=\frac{\[\frac{1}{2}\][2g+2]!}{\[g+{\frac{1}{2}}\]!\ [g]!\ \ [g+1]!}
=\frac{\text{ch}(\mathbf{A})}{\text{ch}(\mathbf{F})}.
\end{align}
Thus from (\ref{ge}) we conclude
$$
\text{ch}(\mathbf{A}_0)=
\frac{\text{ch}(\mathbf{A})}{\text{ch}(\mathbf{F})}
$$
which completes the proof of Proposition 1.
QED.
\section{Appendix B. Proof of Proposition 5.}
We define $W^k$ by the LHS of (\ref{W}):
$$
W^k=
H^k (X_{\text{aff}}(g),\mathbb{C})/
\big(\omega\wedge  H^{k-2}(X_{\text{aff}}(g),\mathbb{C})\big).
$$
We first show that the map
$$
i':H^k(X_{\text{aff}}(g),\mathbb{C}) \lar
{H}^k(X(g)-D,\mathbb{C})
$$
satisfies
$$
i'\big(\omega\wedge  H^{k-2}(X_{\text{aff}}(g),\mathbb{C})\big)=0
$$
and thereby it induces the map
$$
i':W^k\lar {H}^k(X(g)-D,\mathbb{C}).
$$
Next we shall construct a subspace $W_k$ of the homology group
$H_k(X(g)-D,\mathbb{C})$ such that 
the pairing between $W_k$ and $i'(W^k)$ is non-degenerate.
This proves Proposition 5.

Let us study the properties of the differential form $\omega$
defined in (\ref{omega}).
Consider some differentials $\lambda _j$ from 
$H^1(X_{\text{aff}},\mathbb{C})$,
$j=1,\cdots ,k-2$, and construct the $g$-form:
\begin{align}
&\Omega = d\ \(\tilde{\kappa }\wedge
\tilde{\lambda} _1  \wedge \cdots\wedge
\tilde{\lambda} _{k-2}   \) \label{xii}
\end{align} 
where the one form $\tilde{\kappa }$ is given by
\begin{align}
\tilde{\kappa }&=\sum _{i<j}  \kappa ^{(ij)},
\qquad
\kappa ^{(ij)}=
\textstyle{\frac{1}{4}}\frac{y_i-y_j}{z_i-z_j}
\(\frac{dz_i}{y_i}+\frac{dz_j}{y_j} \)
\non
\end{align}
The form under $d$ in RHS of (\ref{xii}) belongs to 
$\mathbf{C}_f^{k-1}$, that is, it has singularity on the divisor
$D=D_0\cup D_\infty$,
but after the differential is applied the singularities
on $D_0$ disappear. 
Indeed, one easily shows that
\begin{align}
&
d\kappa ^{(ij)}=
\textstyle{\frac{1}{4}}\sum\limits _{k=1}^g (2k-1)
\(\nu _k^{(j)}\nu _{-k+1}^{(i)}-\nu _k^{(i)}\nu _{-k+1}^{(j)}\),
\label{xiii}
\end{align}
where $\nu _j$ are defined as 
$$
\nu _j=q_j(z)\frac{dz}{y},\qquad j=-(g-1),\cdots ,g-1,g,
$$
and $q_j$ is the following polynomial of degree  $g-j$:
$$
q_j(z)=
\text{res}_{p_1=\infty}\(\frac{y_1 z_1^{-j}}{z_1-z }d\sqrt{z_1}\).
$$
The differentials $\nu _j$ are normalized at
infinity as
$$
\nu _j\sim
-2\( z^{-j}+O(z^{-g-1})\)
d\sqrt{z}\qquad \text{for}
\qquad p\to\infty.
$$
The differentials $\nu _j$ for $j=1,\cdots ,g$ are holomorphic and
$\nu _j$ for $j=-g+1,\cdots ,0$ are of the second kind.
It is easy to verify that
$$
\nu _i\circ\nu _j=\frac{4}{j-i}\  \delta _{i+j,1}.
$$
This means, in particular, that, for two cycles $\gamma_1$, $\gamma_2$
on $X$
\begin{eqnarray}
\int_{\gamma_1\times\gamma_2}d\kappa^{(12)}
&=&
\textstyle{\frac{1}{4}}\sum\limits _{k=1}^g (2k-1)
\(
\int_{\gamma_1}\nu _{-k+1}\int_{\gamma_2}\nu _k
-
\int_{\gamma_2}\nu _{-k+1}\int_{\gamma_1}\nu _k
\)
\nonumber
\\
&=&
\gamma_1\circ\gamma_2,
\label{intersectionno}
\end{eqnarray}
due to Riemann bilinear relation, where $\gamma_1\circ\gamma_2$
is the intersection number.
This is an important property of $d\kappa^{(ij)}$.

The equation (\ref{xiii}) means that
\begin{align}
&\Omega = d\ \(\tilde{\kappa }\wedge
\tilde{\lambda} _1  \wedge \cdots\wedge
\tilde{\lambda} _{k-2}   \) 
=
-\omega\wedge
\tilde{\lambda} _1  \wedge \cdots\wedge
\tilde{\lambda} _{k-2}\non
\end{align} 
where we have to remind that
$$
\omega =\textstyle{\frac{1}{4}}\sum\limits _{k=1}^g (2k-1)
\ \tilde{\nu} _k\wedge \tilde{\nu} _{-k+1}.
$$
This proves that 
$i'\Big(\omega\wedge H^{k-2}(X_{\text{aff}}(g),\mathbb{C})\Big)=0.$


Consider the homology groups of $\S_{\text{aff}}(g)$.
Taking dual to the relation (\ref{wedge})
we obtain a similar relation for the homology groups:
\begin{align}
&
H_k(\S_{\text{aff}}(g),\mathbb{C})
\simeq 
\bigwedge^k H_1(\S_{\text{aff}}(g),\mathbb{C}).
\non
\end{align}

The first homology group $H_1(\S_{\text{aff}}(g),\mathbb{C})$
is isomorphic to $H_1(\S_{\text{aff}},\mathbb{C})$. To have an
element $\tilde{\delta}$ from $H_1(\S_{\text{aff}}(g),\mathbb{C})$
one takes a cycle $\delta$ from $H_1(\S_{\text{aff}},\mathbb{C})$ and
symmetrizes it over $g$ copies of $\S_{\text{aff}}$.
In other words, if we fix a point $p_0$ in $X_{\text{aff}}$,
then
\begin{eqnarray}
&&
\tilde{\delta}=\delta^{(1)}+\cdots+\delta^{(g)},
\nonumber
\\
&&
\delta^{(i)}=p_0\otimes\cdots\otimes\delta\otimes\cdots\otimes p_0
\in 
H_0^{\otimes (i-1)}\otimes H_1\otimes H_0^{\otimes (g-i)}
\hookrightarrow
H_1(\S_{\text{aff}}^g,\mathbb{C}),
\nonumber
\end{eqnarray} 
where $H_j=H_j(\S_{\text{aff}},\mathbb{C})$.
 
There is an obvious embedding
$
\S(g)-D
$
into  
$\S_{\text{aff}}(g)$.
It induces a map between the homology groups
\begin{align}
&H_k(\S(g)-
D,\mathbb{C})\ \stackrel{i}{\longrightarrow }
\  H_k(\S_{\text{aff}}(g),\mathbb{C}). 
\label{v}
\end{align}
The meaning of this map is simple: every cycle on 
$
\S(g)-
D
$
is at the same time a cycle on
$\S_{\text{aff}}(g)$.
There are two subtleties:\newline
1. Nontrivial cycle on  
$
\S(g)-
D
$
can be trivial on 
$\S_{\text{aff}}(g)$
i.e. the map (\ref{v}) can have kernel.\newline
2. On 
$\S_{\text{aff}}(g)$
there are cycles that intersect with
$D_0$ which means that they are not cycles on 
$
\S(g)-
D
$, so, the
map (\ref{v}) can have cokernel.

Let us study the image of the map (\ref{v}). 
A $k$-cycle from $H_k(\S_{\text{aff}}(g),\mathbb{Z})$
is a linear combination of elements of the form:
$$
\Delta=\tilde{\delta}_1\wedge\cdots\wedge\tilde{\delta}_k,
$$
where $\delta_j\in H_1(\S_{\text{aff}},\mathbb{Z})$.
The product 
$$
\Delta^\prime=
\delta_1\times\cdots\times\delta_k\times p_0\times\cdots\times p_0
$$
defines an element of $H_k(\S_{\text{aff}}^g,\mathbb{Z})$.
Let $\pi$ be the projection map 
$X_{\text{aff}}^g\lar X_{\text{aff}}(g)$ and 
$\pi_\ast$ the induced map on the homology groups,
$\pi_\ast:H_k(X_{\text{aff}}^g,\mathbb{Z})
\lar H_k(X_{\text{aff}}(g),\mathbb{Z})$.
Then $\Delta=k!{\textstyle\binom{g}{k}}\pi_\ast(\Delta^\prime)$
in $H_k(X_{\text{aff}}(g),\mathbb{Z})$.

Thus the cycle $\Delta$ belongs to $\text{Im}(i)$
if $\Delta^\prime$ does not intersect $\pi^{-1}(D)$. 
Recall that $D=D_0\cup D_{\infty}$. 
By construction $\Delta^\prime$ has no intersection with
$\pi^{-1}(D_{\infty})$. 
One easily realizes that $\Delta^\prime$
does not
intersect with $\pi^{-1}(D_0)$ iff 
$$
\delta_i\circ \sigma(\delta_j)=0
\qquad 
\forall \ i,j.
$$

It is rather obvious property of the hyper-elliptic
involution that
$$ 
\delta_i\circ \sigma(\delta_j)
= -\delta_i\circ \delta_j.
$$
Hence we come to the following 
\vskip 3mm
\noindent
{\bf Proposition B.}
{\it The} $\text{Im}(i)$ {\it  contains linear
combination of cycles }
$$
\Delta=\tilde{\delta}_1\wedge\cdots\wedge\tilde{\delta}_k,
\quad
\delta_1,\cdots,\delta_k\in H_1(X_{\text{aff}},\mathbb{Z}),
$$
{\it such that
$$
\delta_i\circ \delta_j =0
\qquad 
\forall \ i,j.
$$ }
\vskip 3mm

\noindent
Let $W_k$ denote the space obtained as $\mathbb{C}$-linear
span of cycles in $H_k(X(g)-D,\mathbb{Z})$ corresponding to 
$\Delta$'s in this proposition.

Take a canonical cycles $\alpha_i$, $\beta_j$ and set
$$
A_i=\tilde{\alpha}_i,
\quad
A_{i+g}=\tilde{\beta}_i,
\quad
1\leq i\leq g.
$$
Define
\begin{eqnarray}
&&
V_{\mathbb{Z}}=H_1(X_{\text{aff}},\mathbb{Z})=\oplus_{i=1}^{2g}\mathbb{Z}A_i,
\quad
V_{\mathbb{C}}=H_1(X_{\text{aff}},\mathbb{C})=\oplus_{i=1}^{2g}\mathbb{C}A_i.
\nonumber
\end{eqnarray}
The symplectic form on $V_{\mathbb{C}}$ is defined by
$$
A_i\circ A_{j}=\pm\delta_{j,i\pm g}.
$$
By definition 
$$
i(W_k)=\text{Span}_{\mathbb{C}}(U_k),
\quad
U_k=\{\gamma_1\wedge\cdots\wedge\gamma_k\vert \gamma_1,\cdots,\gamma_k\in 
V_{\mathbb{Z}},\  \gamma_i\circ \gamma_j=0\  \forall i,j\}.
$$
Define
$$
\tilde{W}_k=\text{Span}_{\mathbb{C}}(\tilde{U}_k),
\quad
\tilde{U}_k=
\{\gamma_1\wedge\cdots\wedge\gamma_k\vert \gamma_1,\cdots,\gamma_k\in 
V_{\mathbb{C}},\  \gamma_i\circ \gamma_j=0\  \forall i,j\}.
$$
Consider the map
\begin{eqnarray}
\varphi_k&:&\wedge^kV_{\mathbb{C}}\lar \wedge^{k-2}V_{\mathbb{C}},
\nonumber
\\
\varphi_k&&(\gamma_1\wedge\cdots\wedge\gamma_k)=
\sum_{i<j}(-1)^{i+j-1}(\gamma_i\circ \gamma_j)\gamma_{\{ij\}},
\label{sl2e}
\end{eqnarray}
where $\gamma_{\{ij\}}$ is obtained from 
$\gamma_1\wedge\cdots\wedge\gamma_k$ removing 
$\gamma_i$ and $\gamma_j$.
It is known that, for $k\leq g$, $\varphi_k$ is surjective and its kernel
$\text{Ker}\varphi_k$ is isomorphic to the $k$-th fundamental
irreducible representation of $\text{Sp}(2g,\mathbb{C})$ 
({\it cf.} Theorem 17.5 \cite{FH}).
In particular
\begin{eqnarray}
&&
d_k:=\text{dim}\ \text{Ker}(\varphi_k)=
{\textstyle
\binom{2g}{k}
}
-
{\textstyle
\binom{2g}{k-2}
}.
\nonumber
\end{eqnarray}
The following lemma can be easily proved.
\vskip3mm

\noindent
{\bf Lemma B.}{\it Suppose that $k\leq g$. Then}
$$
i(W_k)=\tilde{W}_k=\text{Ker}(\varphi_k).
$$
\vskip3mm

As a consequence of the lemma one has in particular
\begin{align}
\text{dim}\, W_k
\geq
\text{dim}\ i(W_k)
={\textstyle\binom{2g}{k}}-{\textstyle\binom{2g}{k-2}}.
\label{vi}
\end{align}

Let us show that $W_k$ and $i'(W^k)$ pairs completely.
The pairings
\begin{eqnarray}
&&
<\quad,\quad>_1:
\ H_k(X(g)-D,\mathbb{C})
\otimes
H^k(X(g)-D,\mathbb{C})
\lar
\mathbb{C}
\label{nondegpairing1}
\end{eqnarray}
and
\begin{eqnarray}
&&
<\quad,\quad>_2:
\wedge^k H_1(X_{\text{aff}}(g),\mathbb{C})
\otimes
\wedge^k H^1(X_{\text{aff}}(g),\mathbb{C})
\lar
\mathbb{C}
\label{nondegpairing2}
\end{eqnarray}
are related by
\begin{eqnarray}
&&
<\gamma,i'(\eta)>_1=<i(\gamma),\eta>_2,
\nonumber
\end{eqnarray}
for
$\gamma\in H_k(X(g)-D,\mathbb{C})$,
$\eta\in \wedge^k H^1(X_{\text{aff}}(g),\mathbb{C})$.
The pairing (\ref{nondegpairing2}) is given by the integral:
\begin{eqnarray}
&&
<\tilde{\gamma}_1\wedge\cdots\wedge\tilde{\gamma}_k,
\tilde{\eta}_1\wedge\cdots\wedge\tilde{\eta}_k>_2
\nonumber
\\
&=&
\int_{\tilde{\gamma}_1\wedge\cdots\wedge\tilde{\gamma}_k}
\tilde{\eta}_1\wedge\cdots\wedge\tilde{\eta}_k
=k!{\textstyle\binom{g}{k}}
\text{det}(\int_{\gamma_i}\eta_j)_{1\leq i,j\leq k}.
\nonumber
\end{eqnarray}
By (\ref{intersectionno}) we have
$$
<i(W_k), \omega\wedge^{k-2} H^1(X_{\text{aff}}(g),\mathbb{C})>_2=0.
$$
Thus the pairing (\ref{nondegpairing1}), (\ref{nondegpairing2}) 
induce pairings
\begin{eqnarray}
&&
<\quad,\quad>_1:\ 
W_k\otimes i'(W^k)\lar\mathbb{C}
\label{nondegpairing3}
\\
&&
<\quad,\quad>_2:\ 
i(W_k)\otimes W^k\lar\mathbb{C}.
\label{nondegpairing4}
\end{eqnarray}
Since (\ref{nondegpairing2}) is non-degenerate and 
$\text{dim}\, i(W_k)=\text{dim}\, W^k$, the pairing (\ref{nondegpairing4})
is non-degenerate.
It easily follows from this that 
the pairing (\ref{nondegpairing3}) is also non-degenerate.
Thus we have proved Proposition 5.
QED.

\vskip3mm
\noindent
{\bf Corollary B.} {\it We have $W_k\simeq i(W_k)$. In particular}
$$
\text{dim}\,W_k={\textstyle\binom{2g}{k}}-{\textstyle\binom{2g}{k-2}}.
$$

\section{Appendix C. The case of generic Abelian variety.}

Let $(J,\Theta)$ be a principally polarized Abelian variety
such that $\Theta$ is non-singular.
Then

\vskip 3mm
\noindent
{\bf Proposition C.} {\it
The dimensions of cohomology groups of $J-\Theta$ are given by}
\begin{eqnarray}
\hbox{dim} H^k(J-\Theta,\mathbb{C})
&=& \binom{2g}{k}-\binom{2g}{k-2}, 
\quad
k\leq g-1,
\nonumber
\\
&=&
\binom{2g}{g}-\binom{2g}{g-2}+g!-\frac{(2g)!}{g!(g+1)!},
\quad
k=g,
\nonumber
\\
&=&0,
\quad
k>g.
\nonumber
\end{eqnarray}
\vskip 3mm
\noindent
{\it Proof.}
Consider the inclusions $\Theta\subset J\subset (X,J)$ and 
the induced homology exact sequence:
\begin{eqnarray}
&&
\cdots\lar
H_k(\Theta,\mathbb{C})
\lar
H_k(J,\mathbb{C})
\lar
H_k(J,\Theta)
\lar
H_{k-1}(\Theta,\mathbb{C})
\lar
\cdots.
\nonumber
\end{eqnarray}
Taking the dual sequence of this and using the Poincare-Lefschetz duality
we get
\begin{eqnarray}
\cdots\lar
H_{k-1}(\Theta,\mathbb{C})
\lar
H_k(J-\Theta,\mathbb{C})
&\lar&
H_k(J,\mathbb{C})
\lar
\nonumber
\\
&\lar&
H_{k-2}(\Theta,\mathbb{C})
\lar\cdots.
\label{longexact1}
\end{eqnarray}
Since $J-\Theta$ is affine 
$$
H_k(J-\Theta,\mathbb{C})=0,
\quad
k>g.
$$
Then we have
$$
H_k(\Theta,\mathbb{C})\simeq H_{k+2}(J,\mathbb{C}),\quad k\geq g,
\quad
H_k(\Theta,\mathbb{C})\simeq H_{k}(J,\mathbb{C}),\quad k\leq g-2.
$$
It is easy to check that, for $k\leq g$, the dual map of
\begin{eqnarray}
&&
H_k(J,\mathbb{C})
\lar
H_{k-2}(\Theta,\mathbb{C})
\nonumber
\end{eqnarray}
is given by wedging the fundamental class $[\Theta]$ of $\Theta$:
\begin{eqnarray}
&&
[\Theta]\wedge:
H^{k-2}(\Theta,\mathbb{C})\simeq H^{k-2}(J,\mathbb{C})
\lar
H^k(J,\mathbb{C}).
\label{omegawedge}
\end{eqnarray}
Using the representation theory of $sl_2$ as in the proof of the
hard Lefschetz theorem ({\it c.f.}\cite{GH}), 
the map (\ref{omegawedge}) is injective
for $k\leq g$.
Thus by (\ref{longexact1}) the following exact sequences hold:
\begin{eqnarray}
&&
0\ra H^{k-2}(\Theta,\mathbb{C}) 
\stackrel{[\Theta]\wedge}{\lar} H^{k}(J,\mathbb{C})
\ra H^{k}(J-\Theta,\mathbb{C})
\ra 0,
\quad
k<g,
\nonumber
\\
&&
0\ra H^{g-2}(\Theta,\mathbb{C}) 
\stackrel{[\Theta]\wedge}{\lar} H^{g}(J,\mathbb{C})
\ra H^{g}(J-\Theta,\mathbb{C})
\ra 
\nonumber
\\
&&
\ra H^{g-1}(\Theta,\mathbb{C})
\ra H^{g+1}(J,\mathbb{C})
\ra 0.
\nonumber
\end{eqnarray}
Proposition C follows from these exact sequences and the fact
$$
\chi(\Theta)=(-1)^{g-1}g!.
$$
QED.
\vskip3mm
By Proposition F 3
the fundamental class of $\Theta$ coincides
with $\omega$ in Proposition 5 in the hyper-elliptic case.
Thus, if we define $W^k$ in a similar formula to (\ref{W}),
we have
\begin{eqnarray}
&&
W^k\simeq H^{k}(J-\Theta,\mathbb{C}),
\quad 
k\leq g-1,
\nonumber
\\
&&
W^g \hookrightarrow H^{g}(J-\Theta,\mathbb{C}).
\nonumber
\end{eqnarray}

\section{Appendix D. The proof of Conjecture 3 for $k=1$}
Notice that
$$
H^k(X_{\text{aff}}(g),\mbc)\simeq \wedge^k H^1(X,\mbc)\simeq \wedge^k H^1(J(X),\mbc),
$$
and $X(g)-D\simeq J(X)-\Theta$.
In particular $W^1 \simeq H^1(J(X),\mbc)$.

\vskip 3mm
\noindent
{\bf Proposition D.} {\it
For any principally polarized Abelian variety $(J,\Theta)$ such that
$\Theta$ is irreducible we have the isomorphism
$$
H^1(J,\mbc) \simeq H^1(J-\Theta,\mbc).
$$
}
\vskip2mm
\noindent
{\it Proof.}
Following \cite{AH} we shall use the following notations:
\begin{description}
\item[]
${\cal O}(n\Theta):$
the sheaf of meromorphic functions on 
$J$ which have poles only on $\Theta$ of order at most $n$,
\item[]
${\cal O}(\ast\Theta):$
the sheaf of meromorphic functions on 
$J$ which have poles only on $\Theta$,
\item[]
$\Omega^k(n\Theta):$
the sheaf of meromorphic $k$-forms on 
$J$ which have poles only on $\Theta$ of order at most $n$,
\item[]
$\Omega^k(\ast\Theta):$
the sheaf of meromorphic $k$-forms on 
$J$ which have poles only on $\Theta$,
\item[]
$\Phi^k(n\Theta):$
the sheaf of closed meromorphic $k$-forms on 
$J$ which have poles only on $\Theta$ of order at most $n$,
\item[]
$\Phi^k(\ast\Theta):$
the sheaf of closed meromorphic $k$-forms on 
$J$ which have poles only on $\Theta$,
\item[]
$
R^k(n\Theta)=\Phi^k(n\Theta)/d\big(\Omega^{k-1}((n-1)\Theta)\big),
\quad
R^k(\ast\Theta)=\Phi^k(\ast\Theta)/d\big(\Omega^{k-1}(\ast\Theta)\big).
$
\end{description}
In particular
$$
\Omega^0(n\Theta)={\cal O}(n\Theta),
\quad
\Omega^0(\ast\Theta)={\cal O}(\ast\Theta).
$$

We first recall the description of $H^1(J,\mbc)$ in terms of the
differentials of the first and second kinds.
Consider the sheaf exact sequence:
\begin{eqnarray}
&&
0 \lar \mbc \lar {\cal O}(\ast\Theta) \stackrel{d}{\lar} 
d\big( {\cal O}(\ast\Theta) \big) \lar 0.
\nonumber
\end{eqnarray}
Since 
$$
H^k(J,{\cal O}(\ast\Theta))=0,
\quad
k\geq 1,
$$
we have
\begin{eqnarray}
&&
H^1(J,\mbc)\simeq 
H^0(J,d{\cal O}(\ast\Theta))/d H^0(J,{\cal O}(\ast\Theta)).
\label{secondkind}
\end{eqnarray}
The numerator in the right hand side of (\ref{secondkind}) is nothing but
the space of differential one forms of 
the first and the second kinds on $J$ and 
the denominator is the space of globally exact meromorphic one forms.

On the other hand, by the algebraic de Rham theorem,
the first cohomology group of the affine variety $J-\Theta$ is
described as
\begin{eqnarray}
&&
H^1(J-\Theta,\mbc)\simeq
H^0(J,\Phi^1(\ast\Theta))/d H^0(J,{\cal O}(\ast\Theta)).
\label{deRham}
\end{eqnarray}

Comparing (\ref{secondkind}) and (\ref{deRham}) what we have to prove is
\begin{eqnarray}
&&
H^0(J,d{\cal O}(\ast\Theta))\simeq H^0(J,\Phi^1(\ast\Theta)).
\label{shoudprove}
\end{eqnarray}

Consider the exact sequence 
$$
0 \lar d{\cal O}(\ast\Theta) \lar \Phi^1(\ast\Theta) \lar R^1(\ast\Theta)
\lar 0.
$$
The cohomology sequence of this is
$$
0\lar H^0(J,d{\cal O}(\ast\Theta))
\lar
H^0(\Phi^1(\ast\Theta))
\lar
H^0(R^1(\ast\Theta))
\lar\cdots.
$$
{}From this what should be proved is that the map
\begin{eqnarray}
&&
H^0(\Phi^1(\ast\Theta))
\lar
H^0(R^1(\ast\Theta))
\nonumber
\end{eqnarray}
is a $0$-map.
To study this map we refer the lemma from \cite{AH}.

\vskip3mm
\noindent
{\bf Lemma D.}(Lemma 8 \cite{AH}) {\it
\begin{description}
\item[(1)] $R^1(\ast\Theta)\simeq\mbc_\Theta$, where $\mbc_\Theta$
is the constant sheaf on $\Theta$ and the isomorphism is given by
$$
[\frac{d\theta}{\theta}]\llar [1_\Theta]
$$
at any stalk.
\item[(2)] $R^1(\ast\Theta)\simeq R^1(n\Theta)$, $n=1,2,\cdots$.
\end{description}
}
\vskip5mm

In the proof of Lemma D (1) we use our assumption that 
$\Theta$ is irreducible.

Using this lemma we reduce the problem from "$\ast\Theta$" to 
"$n\Theta$" with finite $n$.

Consider the exact sequence
\begin{align}
&
0\lar d{\cal O}(n\Theta) \lar \Phi^1((n+1)\Theta) 
\lar R^1(\ast\Theta)\lar 0,
\quad
n=0,1,\cdots.
\label{residue}
\end{align}
{From} the cohomology sequence of it we have the map
$$
H^0(J,R^1(\ast\Theta))\lar H^1(J,d{\cal O}(n\Theta))
$$
which we denote by $\pi_n$.
Let us prove
$$
\text{Ker}\pi_n=0,
\quad
n=0,1,2,\cdots.
$$
To this end we study $H^1(J,d{\cal O}(n\Theta))$.
Using the exact sequence
$$
0\lar \mbc \lar {\cal O}(n\Theta) \stackrel{d}{\lar} 
d{\cal O}(n\Theta) \lar 0,
$$ 
we easily have
\begin{align}
&
H^1(J,d{\cal O}(n\Theta))\simeq H^2(J,\mbc),
\quad n\geq 1,
\label{h11}
\\
&
H^1(J,d{\cal O})\hookrightarrow H^2(J,\mbc).
\label{h12}
\end{align}
The natural maps 
$$
d{\cal O}(n\Theta)\lar d{\cal O}((n+1)\Theta),
\ \Phi^1(n\Theta)\lar \Phi^1((n+1)\Theta),
\ R^1(n\Theta)\simeq R^1((n+1)\Theta),
$$ 
and the sequence (\ref{residue}) induce a commutative
diagram of cohomology groups.
It follows from this commutative diagram and (\ref{h11}), (\ref{h12})
that $\text{Ker}\pi_0=0$ implies $\text{Ker}\pi_n=0$ for $n\geq 1$.
Now $\text{Ker}\pi_0=0$ follows from 
$$
H^0(J, d{\cal O})\simeq H^0(J, \Omega^1),
\quad
H^0(J,\Phi^1(\Theta))\simeq H^0(J,\Omega^1).
$$
The second isomorphism follows from the fact that a 
meromorphic function on $J$ which has poles only on $\Theta$
of order at most one is a constant.
Thus the Proposition D is proved.
QED.

\section{Appendix E.}
Recall that, for a set of complex numbers $f^0=(f_1^0,\cdots,f_{2g+1}^0)$,
the ring $\mathbf{A}_{f^0}$ is defined by
\begin{eqnarray}
&&
\mathbf{A}_{f^0}=\mathbf{A}\otimes_{\mathbf{F}}\mathbb{C}_{f^0}
=\frac{\mathbf{A}}{\sum_{i=1}^{2g+1}\mathbf{A}(f_i-f_i^0)},
\nonumber
\end{eqnarray}
where $f_i$ is the coefficient of $f(z)$.
If all $f_i^0=0$, then $\mathbf{A}_{f^0}=\mathbf{A}_{0}$.
The ring $\mathbf{A}_{0}$ is graded while $\mathbf{A}_{f^0}$ is not graded
unless all $f_i=0$. 
Instead $\mathbf{A}_{f^0}$ 
is a filtered ring for any $f^0$. Let $\mathbf{A}_{f^0}(n)$ 
be the set of elements of 
$\mathbf{A}_{f^0}$ represented by $\oplus_{k\leq n}\mathbf{A}^{(n)}$.
Then
$$
\mathbf{A}_{f^0}=\cup_{n\geq 0}\mathbf{A}_{f^0}(n),
\quad
\mathbf{A}_{f^0}(0)=\mathbb{C}
\subset
\mathbf{A}_{f^0}\(\textstyle{\frac{1}{2}}\)
\subset
\mathbf{A}_{f^0}(1)
\subset\cdots.
$$
Consider the graded ring associated with this filtration:
$$
\text{gr}\mathbf{A}_{f^0}=
\oplus \text{gr}_n\mathbf{A}_{f^0},
\quad
\text{gr}_n\mathbf{A}_{f^0}=
\frac{\mathbf{A}_{f^0}(n)}{\mathbf{A}_{f^0}(n-\frac{1}{2})}.
$$
Since $\text{deg}(f^0_i)=0$, $\text{gr}\mathbf{A}_{f^0}$ becomes
a quotient of $\mathbf{A}_0$.
In other words there is a surjective ring homomorphism
\begin{eqnarray}
&&
\mathbf{A}_0\lar \text{gr}\mathbf{A}_{f^0}.
\label{gradedmap}
\end{eqnarray}
We shall prove that this map is injective.

\vskip3mm
\noindent
{\bf Proposition E.} {\it There is an isomorphism of graded rings:}
$$
\mathbf{A}_0\simeq \text{gr}\mathbf{A}_{f^0}.
$$
\vskip3mm
\noindent
{\it Proof.} Notice that the map (\ref{gradedmap}) respects the grading and
it is surjective at each grade.
By Proposition 2,
$\mathbf{A}\simeq \mathbf{F}\otimes_{\mathbb{C}} \mathbf{A}_0$ as a
$\mathbb{C}$-vector space.
Thus we have $\mathbf{A}_0\simeq \mathbf{A}_{f^0}$ as a $\mathbb{C}$-vector
space for any $f^0$.
In particular the basis of $\mathbf{A}_{0}$ given in Proposition 1 is
also a basis of $\mathbf{A}_{f^0}$.
Denote by $\{x^{(n)}_j\}$ the basis of the degree $n$ part 
$\mathbf{A}_{0}^{(n)}$.
Let us prove that $\{x^{(n)}_j\}$ are linearly independent in
$\text{gr}\mathbf{A}_{f^0}$ by the induction on $n$.
For $n=0$ the statement is obvious.
We assume that the statement is true for all $m$ satisfying $m<n$.
This means that $\{x^{(m)}_j\}$ is a basis of the degree $m$
part of $\text{gr}\mathbf{A}_{f^0}$ for all $m<n$.
In particular $\{x^{(m)}_j\vert m<n\}$ is a basis of 
$\mathbf{A}_{f^0}(n-\textstyle{\frac{1}{2}})$.
Suppose that the relation 
$$
\sum_j \alpha_jx^{(n)}_j\in \mathbf{A}_{f^0}(n-\textstyle{\frac{1}{2}}).
$$
holds, where some $\alpha_j\neq 0$. Then this means that
$\{x^{(m)}_j \vert m\leq n\}$ are linearly dependent in 
$\mathbf{A}_{f^0}$. This contradicts the fact
that $\{x^{(k)}_j\}$ is a basis of $\mathbf{A}_{f^0}$.
Thus $\{x^{(n)}_j\}$ are linearly independent in $\text{gr}\mathbf{A}_{f^0}$.
QED.
\section{Appendix F. Construction of a free resolution of $\mathbf{A}_0$}
Recall that 
$$
{\cal D}=\mathbb{C}\ [D_1,\cdots,D_g]
$$ 
is the polynomial ring generated by the commuting vector fields
$D_1,\cdots,D_g$.
Then $\mathbf{A}$, $\mathbf{A}_0$ and $\mathbf{A}_{f^0}$ are 
${\cal D}$ modules.
We shall construct a free ${\cal D}$-resolution of $\mathbf{A}_0$
assuming Conjecture 2 and 3.
To avoid the notational confusion we shall describe the space
$W^k$ using the abstract vector space $V$ of dimension $2g$ with
a basis $v_i$, $\xi_i$ ($1\leq i\leq g$):
$$
V=\oplus_{i=1}^g\mathbb{C}v_i\oplus_{i=1}^g\mathbb{C}\xi_i.
$$
Set 
\begin{eqnarray}
&&
\omega=\sum_{i=1}^gv_i\wedge \xi_i\in \wedge^2 V
\nonumber
\end{eqnarray}
and define
\begin{eqnarray}
&&
W^k=\frac{\wedge^k V}{\omega \wedge^{k-2}V}.
\label{abstWk}
\end{eqnarray}
Assign degrees to the basis elements by
\begin{eqnarray}
&&
\hbox{deg}(v_i)=-(i-\textstyle{\frac{1}{2}}),
\quad
\hbox{deg}(\xi_i)=i-\textstyle{\frac{1}{2}}.
\nonumber
\end{eqnarray}
Then $W^k$ defined by (\ref{abstWk}) has the same character
as $W^k$ defined in  Section 6. This justifies the use of the
same symbol.
Consider the free ${\cal D}$-module ${\cal D}\otimes_{\mathbb{C}}W^k$
generated by $W^k$.
We shall construct an exact sequence of ${\cal D}$-modules
of the form
\begin{align}
&
0
\leftarrow
\mathbf{A}_0d\tau_1\wedge\cdots\wedge d\tau_g
\leftarrow
{\cal D}\otimes_{\mathbb{C}}W^g
\stackrel{d}{\leftarrow}
{\cal D}\otimes_{\mathbb{C}}W^{g-1}
\stackrel{d}{\leftarrow}
\cdots
\stackrel{d}{\leftarrow}
{\cal D}\otimes_{\mathbb{C}}W^{0}
\leftarrow 
0.
\nonumber
\end{align}
Define the map
\begin{eqnarray}
&&
{\cal D}\otimes_{\mathbb{C}}\wedge^k V 
\stackrel{d}{\lar}
{\cal D}\otimes_{\mathbb{C}}
\wedge^{k+1}V,
\nonumber
\end{eqnarray}
by
\begin{eqnarray}
&&
d(P\otimes v_I\wedge \xi_J)=
\sum_{i=1}^gD_iP \otimes v_i\wedge v_I\wedge \xi_J,
\nonumber
\end{eqnarray}
where for 
$I=(i_1,\cdots,i_r)$, $v_I=v_{i_1}\wedge\cdots\wedge v_{i_r}$
etc., $P\in {\cal D}$ and $D_iP$ is the product of 
$D_i$ and $P$ in ${\cal D}$.
Since the map $d$ commutes with the map taking 
the wedge with $Q\otimes \omega$ for any $Q\in {\cal D}$:
$$
d(QP\otimes \omega\wedge v_I\wedge \xi_J)
=
\sum_{i=1}^gQD_iP \otimes \omega\wedge v_i\wedge v_I\wedge \xi_J
=(Q\otimes \omega)\wedge d(P\otimes v_I\wedge \xi_J),
$$
it induces a map
$$
{\cal D}\otimes_{\mathbb{C}}W^k 
\stackrel{d}{\lar} 
{\cal D}\otimes_{\mathbb{C}}
W^{k+1}.
$$
It is easy to check that the map $d$ satisfies $d^2=0$.
\vskip5mm

\noindent
{\bf Proposition F 1. }{\it The complex}
\begin{eqnarray}
&&
0
{\llar}
{\cal D}\otimes_{\mathbb{C}}W^g
\stackrel{d}{\llar}
{\cal D}\otimes_{\mathbb{C}}W^{g-1}
\stackrel{d}{\llar}
\cdots
\stackrel{d}{\llar}
{\cal D}\otimes_{\mathbb{C}}W^{0}
\llar
0
\label{conjF}
\end{eqnarray}
{\it is exact at ${\cal D}\otimes_{\mathbb{C}}W^{k}$ except $k=g$.}
\vskip3mm
\noindent
{\it Proof.} 
Notice that the following two facts:
\par
\noindent
1. The following
complex is exact at ${\cal D}\otimes \wedge^k V$ except $k\neq g$:
\begin{eqnarray}
&&
0
\llar
{\cal D}\otimes_{\mathbb{C}}\wedge^gV
\stackrel{d}{\lar}
\cdots
\stackrel{d}{\llar}
{\cal D}\otimes_{\mathbb{C}}V
\stackrel{d}{\llar}
{\cal D}
\llar
0.
\label{Koszul1}
\end{eqnarray}
\par
\noindent
2. The map $\omega\wedge:\wedge^k V\lar \wedge^{k+2} V$ is injective
for $k\leq g-2$.
\par
The property 1 is the well known property of the Koszul complex.
The property 2 is also well known and easily proved using the representation
theory of $sl_2$.

Now suppose that $x\in {\cal D}\otimes_{\mathbb{C}}\wedge^k V$, $k<g$
and $dx=\omega\wedge y$ for some 
$y\in {\cal D}\otimes_{\mathbb{C}}\wedge^{k-2} V$.
Then $\omega\wedge dy=0$ and thus $dy=0$ by the property 2.
Then $y=dz$ for some $z$ by the property 1.
Thus we have $d(x-\omega\wedge z)=0$ and $x=dw+\omega\wedge z$
for some $w$ again by the property 1.
QED.
\vskip2mm

Next we shall define a map from ${\cal D}\otimes_{\mathbb{C}}W^{g}$ 
to $\mathbf{A}_0d\tau_1\wedge\cdots\wedge d\tau_g$.
To this end we need to identify $W^k$ in this section and 
that of Section 6, which is defined as the quotient of 
the cohomology groups of a Jacobi variety.
For this purpose we describe the cohomology groups of
a Jacobi variety in terms of theta functions.

Define 
\begin{eqnarray}
&&
\zeta_i(w)=D_i\log\theta(w)=
\frac{\partial}{\partial \tau_i}\log\theta(w).
\nonumber
\end{eqnarray}
Then, for each $i$, the differential $d\zeta_i(w)$
defines a meromorphic differential form on $J(X)$ which has double poles
on $\Theta$ and which is locally exact.
This means that $d\zeta_i$ is a second kind differential on $J(X)$.
By (\ref{secondkind}) first and second kinds 
differential one forms define elements of $H^1(J(X),\mathbb{C})$. 
The pairing with $H_1(J(X),\mathbb{C})$ is given by integration.
The following proposition can be easily proved by calculating periods.

\vskip5mm

{\bf Proposition F 2.}{\it The first and the second kinds differentials
$d\tau_1$,...,$d\tau_g$, $d\zeta_1$,...,$d\zeta_g$ give a basis of the
cohomology group $H^1(J(X),\mathbb{C})$.}
\vskip3mm

Thus we can identify $V$ with 
$H^1(J(X),\mathbb{C})\simeq H^1(X_{\text{aff}}(g),\mathbb{C})$
by
\begin{eqnarray}
&&
v_i=d\tau_i,
\quad
\xi_i=d\zeta_i.
\nonumber
\end{eqnarray}
The next proposition can be easily proved by calculating
integrals over two cycles in a similar way to \cite{mum}(vol.I, p188).
\vskip5mm

\noindent
{\bf Proposition F 3.}
{\it Let $\omega$ be defined in Proposition 5. Then we have}
$$
\frac{1}{2\pi \sqrt{-1}}\sum_{i=1}^gd\tau_i\wedge d\zeta_i=\omega
$$
{\it as elements in $\wedge^2 H^1(J(X),\mathbb{C})$.
Moreover $\omega$ represents the fundamental class of the theta divisor
$\Theta$ in $H^2(J(X),\mathbb{C})$.}
\vskip3mm

{}From this proposition it is possible to identify $W^k$ in this section and 
that in the previous sections. 
Define the map
\begin{eqnarray}
&&
\wedge^gV \ra \mathbf{A}_{f^0} d\tau_{1}\wedge\cdots\wedge d\tau_{g},
\nonumber
\\
&&
v_I\wedge\xi_J 
\mapsto
d\tau_I\wedge d\zeta_J,
\nonumber
\end{eqnarray}
where for $I=(i_1,\cdots,i_r)$, 
$d\tau_I=d\tau_{i_1}\wedge\cdots\wedge d\tau_{i_r}$ etc.
and $f^0$ is any set of complex numbers such that $y^2=f^0(z)$ 
defines a non-singular curve $X$.

This map extends to the map of ${\cal D}$ modules
\begin{eqnarray}
&&
{\cal D}\otimes_{\mathbb{C}} \wedge^g V
\ra 
\mathbf{A}_{f^0} d\tau_{1}\wedge\cdots\wedge d\tau_{g},
\nonumber
\end{eqnarray}
in the following manner.
Let $P\in {\cal D}$ and consider $P\otimes (v_I\wedge\xi_J)$.
Write $d\zeta_J=\sum_K F_J^Kd\tau _K$, $F_J^K\in \mathbf{A}_{f^0}$.
Then we define 
\begin{eqnarray}
&&
P \otimes (v_I\wedge\xi_J)
\lar
\sum P(F_J^K)d\tau_I\wedge d\tau_K.
\nonumber
\end{eqnarray}

Since, as a meromorphic differential form on $J(X)$,
\begin{eqnarray}
&&
\sum_{i=1}^g d\tau_i\wedge d\zeta_i=
\sum_{i,j=1}^g\frac{\partial^2 \log \theta(w)}{\partial\tau_i\partial\tau_j}
d\tau_i\wedge d\tau_j=0,
\nonumber
\end{eqnarray}
the above map induces the map 
$$
{\cal D}\otimes_{\mathbb{C}} W^g \stackrel{ev}{\lar} 
\mathbf{A}_{f^0} d\tau_{1}\wedge\cdots\wedge d\tau_{g}.
$$
Denote by $({\cal D}\otimes_{\mathbb{C}} W^g)_n$ the subspace of elements
with degree $n$ and by $ev_n$ the restriction of $ev$ to 
$({\cal D}\otimes_{\mathbb{C}} W^g)_n$.
By the definition, for $x\in ({\cal D}\otimes_{\mathbb{C}} W^g)_n$,
$ev_n(x)\in \mathbf{A}_{f^0}(n+g^2/2)$.
By Proposition E, there is a natural isomorphism
$\mathbf{A}_0\simeq \text{gr}\mathbf{A}_{f^0}$ 
(see Appendix E for the filtration of $\mathbf{A}_{f^0}$).
Composing the map $ev_n$ with the natural projection map
$\mathbf{A}_{f^0}(k)\ra \hbox{gr}_k\mathbf{A}_{f^0}\simeq \mathbf{A}_0^{(k)}$ 
we obtain the map, which we denote by $ev_n$ too,
\begin{eqnarray}
&&
({\cal D}\otimes_{\mathbb{C}} W^g )_n
\stackrel{ev_n}{\lar} 
 (\mathbf{A}_0 d\tau _{1}\wedge\cdots\wedge d \tau_{g})_n,
\nonumber
\end{eqnarray}
where the RHS means the degree $n$ subspace.
Taking the sum of $ev_n$ we finally have the map
\begin{eqnarray}
&&
ev=\oplus_n ev_n: {\cal D}\otimes_{\mathbb{C}} W^g 
{\lar} 
\mathbf{A}_0 d\tau _{1}\wedge\cdots\wedge d \tau_{g},
\nonumber
\end{eqnarray}
which we also denote by the symbol $ev$.

If we assume Conjecture 2 and 3, $H^k(\mathbf{C}^\ast_0)\simeq W^k$.
If this holds, then $ev$ is surjective by Proposition 3.
\vskip5mm

\noindent
{\bf Lemma F.}
{\it Suppose that Conjecture 2 and 3 are true.
Then the kernel of $ev$ is given by}
$$
\hbox{Ker}(ev)=d\big({\cal D}\otimes W^{g-1} \big).
$$
\vskip3mm
\noindent
{\it Proof.} It is easy to check that
$$
\hbox{Ker}(ev)\supset d\big({\cal D}\otimes W^{g-1} \big).
$$
Since $ev$ is surjective and 
$$
\text{ch}(\mathbf{A}_0)=\text{ch}\Big(
\frac
{{\cal D}\otimes_{\mathbb{C}}W^g }
{d\big({\cal D}\otimes_{\mathbb{C}}W^{g-1}\big)}
\Big),
$$
the claim of the lemma follows.
QED.

We summarize the result as

\vskip5mm

\noindent
{\bf Theorem F.} 
{\it Suppose that Conjecture 2, 3 are true.
Then the following complex gives a resolution of
$\mathbf{A}_0d\tau_1\wedge\cdots\wedge d\tau_g$ as a ${\cal D}$-module:}
\begin{eqnarray}
&&
0
{\llar}
\mathbf{A}_0d\tau_1\wedge\cdots\wedge d\tau_g
\stackrel{ev}{\llar}
{\cal D}\otimes_{\mathbb{C}}W^g
\stackrel{d}{\llar}
\cdots
\stackrel{d}{\llar}
{\cal D}\otimes_{\mathbb{C}}W^{0}
\llar
0.
\nonumber
\end{eqnarray}
\vskip5mm



\begin{thebibliography}{XXXXXXX}

\bibitem{mum} D. Mumford, {\it Tata Lectures on Theta}, vol. I and II,
Birkh\"auser, Boston (1983)


\bibitem{bbs} O. Babelon, D. Bernard, F.A. Smirnov, Comm. Math. Phys.
{\bf 186} (1997) 601

\bibitem{toda} F.A. Smirnov, J. Phys. A: Math. Gen. {\bf 31} (1998) 8953

\bibitem{skl} E.K. Sklyanin, {\it Separation of Variables},
Progress in Theoretical Physics Supplement {\bf 118} (1995) 35  

\bibitem{Beauv} A. Beauville, Acta Math. {\bf 164} (1990) 211

\bibitem{GH} P. Griffiths and J. Harris, {\it Principles of Algebraic Geometry}, 
A Wiley-Interscience publication (1978)

\bibitem{Mac} I.G. Macdonald, Topology {\bf 1} (1962) 319

\bibitem{AH} M. Atiyah and W.V.D. Hodge, Ann. of Math. {\bf 62} (1955) 56


\bibitem{FH} W. Fulton and J. Harris, {\it Representation Theory,}
Springer, New York (1991)



\end{thebibliography}
\end{document}